\documentclass[journal, onecolumn, 12 pt]{IEEEtranTCOM}
\normalsize

\usepackage{setspace}
\usepackage{cite}
\usepackage{bm}
\usepackage[fleqn]{amsmath}
\usepackage{amsthm}
\usepackage{amssymb}

\usepackage{caption}
\usepackage{subcaption}
\usepackage{graphicx}
\usepackage{dcolumn}
\usepackage{algorithmic}
\usepackage{dsfont}
\usepackage{bbm}
\usepackage{color}
\usepackage[dvipsnames]{xcolor}

\usepackage{algorithm}
\usepackage{multicol}

\newtheorem{proposition}{Proposition}

\theoremstyle{remark}
\newtheorem*{remark}{Remark}

\theoremstyle{OutlineProof}

\theoremstyle{example}

\theoremstyle{example_nc}

\begin{document}
\doublespacing
\title{Robust Clock Skew and Offset Estimation for IEEE 1588 in the Presence of Unexpected {\color{black} Deterministic Path Delay} Asymmetries}
%
%
%

\author{ Anantha K. Karthik, \IEEEmembership{Student Member, IEEE} and Rick S. Blum, \IEEEmembership{IEEE Fellow} 
	
\thanks{This work was supported by the Department of Energy under Award DE-OE0000779.}
	
\thanks{Anantha K. Karthik and Rick S. Blum are with the Department of Electrical and Computer Engineering, Lehigh University, Bethlehem, PA 18015 USA (e-mail: akk314@lehigh.edu; rblum@lehigh.edu).}}

%
%

\markboth{IEEE Transactions on Communications}%
{Submitted paper}
%


\maketitle

\begin{abstract}
IEEE 1588, built on the classical two-way message exchange scheme, is a popular clock synchronization protocol for packet-switched networks. Due to the presence of random queuing delays in a packet-switched network, the joint recovery of the clock skew and offset from the timestamps of the exchanged synchronization packets can be treated as a statistical estimation problem. In this paper, we address the problem of clock skew and offset estimation for IEEE 1588 in the presence of possible unknown asymmetries between the {\color{black} deterministic path delays} of the forward master-to-slave path and reverse slave-to-master path, which can result from incorrect modeling or cyber-attacks. First, we develop lower bounds on the mean square estimation error for a clock skew and offset estimation scheme for IEEE 1588 assuming the availability of multiple master-slave communication paths and complete knowledge of the probability density functions (pdf) describing the random queuing delays. Approximating the pdf of the random queuing delays by a mixture of Gaussian random variables, we then present a robust iterative clock skew and offset estimation scheme that employs the space alternating generalized expectation-maximization (SAGE) algorithm for learning all the unknown parameters. Numerical results indicate that the developed robust scheme exhibits a mean square estimation error close to the lower bounds.  
\end{abstract}



%
\IEEEpeerreviewmaketitle

\section{Introduction}\label{Sec1}
Clock synchronization is a mechanism to ensure a standard reference time across various devices in a distributed network. {\color{black} This is essential for coordinated activities in a network as time provides the only frame of reference between all devices on the network.} Many clock synchronization protocols are available in the literature for synchronizing devices across a packet-switched network. For instance, protocols such as the IEEE 1588 precision time protocol (PTP) \cite{IEEE1588} and the Network Time Protocol (NTP) \cite{NTP} are widely used in IP networks, while protocols such as the Timing Protocol for Sensor Networks (TPSN) \cite{TPSN}, Lightweight Time Synchronization {\color{black} protocol} (LTS) \cite{LTS} and Reference Broadcast Time synchronization (RBS) {\color{black} protocol} \cite{RBS}, are used in wireless sensor networks. In these protocols, the time from a high-cost, high-stability clock (termed master) is distributed to low-cost, low-stability clocks (termed slaves) via an interconnecting network. The clock time at the slave node can be modeled mathematically as a function $c(t)$ of the master node's clock time $t$, i.e.,  $c(t) = \phi t + \delta$ \cite{Noh_2007, Chaudhari_2008, Wu_2011, Anand_2015}, where $\phi$ denotes the relative clock skew and $\delta$ denotes the relative clock offset of the slave's clock time with respect to the master's clock time. If the clocks at the slave and master node are synchronized, then $c(t) = t$. However, in practice, these clocks are not synchronized, implying a synchronization error $e(t) = |c(t) - t|$ that can grow over time. Time synchronization protocols {\color{black} are} utilized to ensure that $e(t)$ remains small.

PTP is a popular clock synchronization protocol used in a number of scenarios, including electrical grid networks \cite{Gaderer_2005}, cellular base station synchronization in 4G Long Term {\color{black} Evolution} (LTE) \cite{Hadzic_2011}, substation communication networks \cite{IEC61850} and industrial control \cite{IEEE_Industrial_Control}. It is cost-effective and offers accuracy comparable to  Global Positioning System (GPS)-based timing. In PTP, as with any clock synchronization protocol built on the two-way message exchange scheme, the slave node exchanges a series of time synchronization packets with the master node and uses the timestamps of these exchanged packets to estimate $\phi$ and $\delta$. The messages traveling between the master and slave nodes can encounter several intermediate switches and routers, accumulating delays at each node. The main factors contributing to the overall delay are: (1) the deterministic (or fixed) propagation and processing delays at the intermediate nodes along the network path between the master and slave nodes and (2) the random queuing delays at each such node. This randomness in the overall network traversal time is referred to as Packet Delay Variation (PDV) \cite{Anand_2015}, and the problem of estimating $\phi$ and $\delta$ in the presence of the PDV is referred to as the ``Clock Skew and Offset Estimation" (CSOE) problem. {\color{black} On the other hand, the problem of estimating $\delta$ in the presence of the PDV when $\phi$ is known is referred to as the ``Clock Offset Estimation" (COE) problem.}

The clock skew and offset can be correctly estimated in PTP (or any time synchronization protocol based on the two-way message exchange scheme such as TPSN and LTS) only if there is a prior known affine relationship between the unknown {\color{black} deterministic path delays} in the forward master-to-slave path and the reverse slave-to-master path \cite{Freris}. However, the presence of an unknown asymmetry between the deterministic path delays can significantly degrade the performance of clock skew and offset estimation schemes \cite{Ullmann_2009}. This unknown asymmetry can arise from several sources, including delay attacks \cite{Ullmann_2009} and routing asymmetry \cite{Balakrishnan}. In this paper, we look to build on our previous works of \cite{AK_Blum_offset, Karthik_PESGM, Karthik_Skew_Offset} to develop {\color{black} joint} clock skew and offset estimation schemes that are robust against unknown {\color{black} deterministic path delay} asymmetries.

{\color{black} Following \cite{Gaderer_2010, Mizrahi_2012, Mizrahi_2012_2, Shpiner_2013}, we assume the availability of multiple master-slave communication paths in our work\footnote{This can be multiple network paths between the master and slave node \cite{Mizrahi_2012, Mizrahi_2012_2, Shpiner_2013} or can be the scenario where the slave node is communicating with multiple master nodes synchronized to the same time \cite{Gaderer_2010}. In our work, we consider the latter scenario of the slave node  communicating with multiple masters synchronized to the same time.}.} We first develop lower bounds on the best possible performance for invariant clock skew and offset estimation schemes in the presence of possible unknown {\color{black} deterministic path delay} asymmetries. Invariant estimators are a reasonable class of estimators in which the estimated parameters scale and shift due to scaling and shifts in the observed timestamps. Also, the commonly employed CSOE schemes in practice are invariant \cite{Karthik_Skew_Offset}. When developing the performance lower bounds, we assume prior knowledge on whether a master-slave communication path has an unknown asymmetry between the {\color{black} deterministic path delays} as well as the complete knowledge of the probability density function (pdf) describing the PDV in the master-slave communication path. The problem of estimating the clock skew and offset in the presence of PDV falls under a variant of the location-scale parameter estimation problems \cite{Berger}, with the unknown clock skew as the scale parameter and the unknown clock offset as the location parameter. Fixing the loss function as the skew-normalized squared error loss, we use invariant decision theory (see Chapter 6 of \cite{Berger}) to design the optimum approach for combining the information from the various master-slave communication paths to estimate the clock skew and offset.  {\color{black} The skew normalized mean square estimation error performance of the developed optimum estimators which assume perfect information (which is usually not available) could be used for comparison against the performance of clock skew and offset estimators that are used in practice.} Further, we show the optimum invariant estimators are minimax optimum, i.e., these optimum estimators minimize the maximum skew normalized mean square estimation error over all parameter values.

In specific scenarios, the complete information regarding the pdf of the PDV might not be readily available. {\color{black} To address this issue, we model the pdf of the PDV by a finite mixture of Gaussian distributions. {\color{black} The Gaussian mixture distribution is a prominent model for approximating a pdf as it is a universal approximator in a certain sense} \cite{Kariya, Blum_spatial, Kostantinos}.} The expectation-maximization (EM) algorithm is a popular iterative approach for obtaining the maximum-likelihood (ML) estimates of the various parameters in problems involving mixture distributions \cite{EM_main_article, EM_mixtures}. {\color{black} To obtain closed-form updates for the various parameters}, we employ the space alternating generalized Expectation-Maximization (SAGE) algorithm \cite{Fessler}, a variant of the EM algorithm, for learning the statistical distribution of the random queuing delays along with the clock skew and offset. The performance of the proposed robust CSOE scheme is evaluated in the LTE backhaul network scenario \cite{Hadzic_2011}. In this scenario, PTP is used to synchronize cellular base station clocks using mobile backhaul networks. Typically, the backhaul networks are leased from a commercial Internet Service Provider (ISP), and the network is shared with other commercial and non-commercial users. The background traffic generated by these users results in PDV for the synchronization packets and traffic models for modeling the background traffic are specified in the ITU-T specification G.8261 standard \cite{ITU}. The class of empirical pdfs corresponding to the PDV according to the ITU-T specification G.8261 \cite{ITU} was derived in \cite{Anand_2015}. In our work, we use the empirical pdfs obtained in \cite{Anand_2015} to evaluate the developed lower bounds as well as the performance of the proposed robust CSOE scheme. Numerical results indicate that the proposed scheme exhibits a skew normalized mean square estimation error close to the performance lower bounds.

{\color{black} 
\subsection{Related work}
In this paper, we have developed a algorithm for jointly estimating the clock skew and offset for PTP (or any clock synchronization protocol based on the two-way message exchange) that is robust against unknown deterministic path delay asymmetries. To the best of our knowledge, there is no such algorithm available in the literature that addresses the problem of jointly estimating clock skew and offset in the presence of unknown deterministic path delay asymmetries. In this section, we present several approaches available in the literature for synchronizing devices in a network, none of which address the problem of jointly estimating clock skew and offset in the presence of unknown deterministic path delay asymmetries. Data Center Time Protocol (DTP) is a clock synchronization protocol that does not use packets, but rather uses the physical layer of network devices to implement a decentralized clock synchronization protocol \cite{DTP_2016}. DTP, however, requires additional hardware, necessitating a fully ``DTP-enabled network" for its deployment \cite{Geng_2018}. HUYGENS \cite{Geng_2018} is a software-based clock synchronization system that incorporates Support Vector Machines and was proposed as an alternative to DTP for synchronizing devices in a data center. In \cite{Qiu_2018}, the authors proposed R-Sync, a time synchronization scheme for the industrial Internet of things (IIoT) as an alternative to existing time synchronization algorithms such as TPSN. Although these protocols offer interesting alternatives to PTP, in our work, we primarily look at improving the performance of PTP (or any clock synchronization protocol based on the two-way message exchange) in the presence of PDV and possible unknown asymmetries between the deterministic path delays in the forward and reverse paths.

We now briefly describe some of the popular algorithms available in the literature that have been developed for clock synchronization protocols built on the two-way message exchange scheme and assume a known relationship between the deterministic path delays of the forward and reverse paths \cite{Noh_2007, Chaudhari_2008, Chaudhari_MVUE_Info, Li_Jeske_2009, PTP_LP, Anand_2015, Anand_2015_Lest}. The ML estimate and the corresponding Cramer-Rao bound (CRB) for the CSOE problem under the Gaussian PDV pdf delay model were derived in \cite{Noh_2007}. Under the exponential PDV pdf delay model, the ML estimate of the clock skew and offset was derived in \cite{Chaudhari_2008, Li_Jeske_2009} and the minimum variance unbiased estimate (MVUE) for the COE problem was derived in \cite{Chaudhari_MVUE_Info}. CSOE schemes for PTP that incorporate linear programming were proposed in \cite{PTP_LP}. For an arbitrary pdf-delay model of PDV, the optimum invariant estimator for the COE problem and CSOE problem was derived in \cite{Anand_2015} and \cite{Karthik_Skew_Offset}, respectively. Further, in \cite{Anand_2015_Lest}, the authors developed a sub-optimal clock offset estimation scheme based on the linear combination of order statistics that exhibits performance close to the performance of the optimum clock offset estimator developed in \cite{Anand_2015}.

We now describe some of the algorithms available in the literature for clock synchronization protocols based on the two-way message exchange scheme which do not assume a prior known relationship between the deterministic path delays of the forward and reverse paths \cite{Sun_2006, Gaderer_2010, Mizrahi_2012, Mizrahi_2012_2, AK_Blum_offset, Karthik_PESGM}. In \cite{Sun_2006}, the authors proposed a COE scheme that uses the median of the observed timing offsets from different master nodes to estimate the clock offset. The proposed scheme is robust against unknown deterministic path delay asymmetries, however, there is a loss in performance due to the significant amount of information being discarded.  In \cite{Gaderer_2010}, the authors proposed the idea of using a group of masters rather than a single master for synchronizing the slave node. The proposed COE scheme works in the presence of a master node failure or a master-slave communication path having unknown deterministic path delay asymmetries. However, it requires prior information regarding the number of the master-slave communication paths that have unknown deterministic path delay asymmetries. This information might not be readily available in many scenarios. Mizrahi \cite{Mizrahi_2012, Mizrahi_2012_2} proposed the use of multiple master-slave communication paths to improve the accuracy of clock offset estimation schemes assuming clock skew is known and also to help protect against delay attacks (a particular case of unknown deterministic path delay asymmetries). Previously, assuming complete knowledge of the clock skew, we developed performance lower bounds along with robust clock offset estimation schemes for PTP that can handle unknown deterministic path delay asymmetries \cite{AK_Blum_offset, Karthik_PESGM}. However, to the best of our knowledge, there is no algorithm available in the literature for the joint estimation of the clock skew and offset that can handle unknown deterministic path delay asymmetries. None of the algorithms presented in \cite{Noh_2007, Chaudhari_2008, Chaudhari_MVUE, Chaudhari_MVUE_Info, Wu_2011, Li_Jeske_2009, PTP_LP, Anand_2015, Anand_2015_Lest, Sun_2006, Gaderer_2010, Mizrahi_2012, Mizrahi_2012_2, AK_Blum_offset, Karthik_PESGM} address the problem of jointly estimating the clock skew and offset for PTP in the presence possible unknown deterministic path delay asymmetries.
}

{\color{black} \subsection{Notations used}
We use bold upper case, bold lower case, and italic lettering to denote matrices, column vectors and scalars respectively. The notations $(.)^T$ and $\otimes$ denote the transpose and Kronecker product, respectively. $\bm{I}_N$ stands for a $N$-dimensional identity matrix, $\mathbbm{1}_N$ denotes a column vector of length $N$ with all the elements equal to $1$ and $\bm{0}_N$ denotes a column vector of length $N$ with all the elements equal to $0$. Further, $\mathbb{R}$ denotes the set of real numbers, $\mathbb{R}^+$ denotes the set of positive real numbers, $\mathbb{R}^+_0$ denotes the set of non-negative real numbers and $\mathcal{I}_A(x)$ denotes the indicator function having the value $1$ when $x \in A$ and $0$ when $x \notin A$.}

\section{Signal Model and Problem Statement}\label{Sec2}
In this section, we briefly describe the two-way message exchange scheme used in PTP and present the considered problem statement along with the assumptions. We assume the availability of $N$ master-slave communication paths and perfect synchronization between the clocks of the $N$ masters. Recall that the relative clock skew and offset of the slave node with respect to a master node are denoted by $\phi \in \mathbb{R}^+$ and $\delta \in \mathbb{R}$, respectively. Assume a total of $P$ rounds of two-way message exchanges at each master-slave communication path. The following sequence of messages are exchanged over the $i^{\mbox{th}}$ master-slave communication path during the $j^{\mbox{th}}$ round of message exchanges: The master node initiates the exchange by sending a \emph{sync} packet to the slave at time $t_{1ij}$. The value of $t_{1ij}$ is later communicated to the slave via a \emph{follow\_up} message. The slave node records the time of reception of the sync message as $t_{2ij}$. The slave node sends a \emph{delay\_req} message to the master node recording the time of transmission as $t_{3ij}$. The master records the time of arrival of the \emph{delay\_req} packet at time $t_{4ij}$ and this value is later communicated to the slave using a \emph{delay\_resp} packet. The relationship between the received timestamps are given by  \cite{Noh_2007, Chaudhari_2008, Chaudhari_MVUE_Info, Wu_2011}
\begin{align}
t_{2ij} = (t_{1ij} + d_i^{ms} + w_{1ij})\phi + \delta & ; &
t_{3ij} =  (t_{4ij} - d_i^{sm} - w_{2ij})\phi + \delta, \label{timestamps}
\end{align}
for $i = 1, 2, \cdots, N$ and $j = 1, 2, \cdots, P$. In (\ref{timestamps}), $d_i^{ms}$ and $d_i^{sm}$ denote the unknown deterministic path delays in the forward and reverse path,  respectively, at the $i^{\mbox{th}}$ master-slave communication path. The variables $w_{1ij}$ and $w_{2ij}$ denote the random queuing delays in the forward and reverse path, respectively, {\color{black} during the $j^{\mbox{th}}$ round of message exchanges for} the $i^{\mbox{th}}$ master-slave communication path. The pdf of $\{w_{kij}\}^P_{j=1}$ is denoted by $f_{ki}(.)$ for $k = 1, 2$ and $i = 1, 2, \cdots, N$.

Freris \emph{et al.} \cite{Freris} provided the necessary conditions for obtaining a unique solution to the clock skew and offset for protocols based on a two-way message exchange scheme {\color{black} for a single forward-reverse path pair of timestamps}. We need to know either one of the {\color{black} deterministic path delays} (either the deterministic delay of the forward path or the deterministic delay of the reverse path), or have a prior known affine relationship between the {\color{black} unknown deterministic path delays} (see Theorem 4 in \cite{Freris}). Synchronization protocols including PTP \cite{IEEE1588}, NTP \cite{NTP}, TPSN \cite{TPSN} used in real networks generally assume that the deterministic path delays in the forward and reverse paths are equal. In this paper, we classify a master-slave communication path as being symmetric or asymmetric depending on the relationship between the deterministic path delays. A \textit{symmetric master-slave communication path} denotes a master-slave communication path in which the {\color{black} deterministic path delays in the forward and reverse paths are equal}, i.e., $d^{ms}_i = d^{sm}_i = d_i$, where $d_i$ denotes the unknown deterministic path delay {\color{black} over} the $i^{\mbox{th}}$ master-slave communication path. Similarly, an \textit{asymmetric master-slave communication path} denotes a master-slave communication path having an unknown asymmetry between the {\color{black} deterministic path delays in the forward and reverse paths,} i.e., $d^{ms}_i = (d_i + \tau_i)$ and $d^{sm}_i = d_i$. The parameter $\tau_i$ denotes the constant {\color{black} (for all $j$) }unexpected asymmetry between the deterministic path delays.

Define $\bm{w}_{ki} = [w_{ki1}, w_{ki2}, \cdots, w_{kiP}]^T$ for $k = 1, 2$, $i = 1, 2, \cdots, N$ and $\bm{t}_{ki} = [t_{ki1}, t_{ki2}, \cdots,$ $t_{kiP}]^T$ for $k = 1, 2, 3, 4$, $i = 1, 2, \cdots, N$.  We now introduce a new binary state vector variable $\bm{\eta} = [\eta_1, \eta_2, \cdots, \eta_N]$, which indicates whether a master-slave communication path is symmetric or asymmetric. The $i^{\mbox{th}}$ element of $\bm{\eta}$ is $1$ when the $i^{\mbox{th}}$ master-slave communication path is asymmetric, else it has a value of $0$. If $\eta_i = 0$, the received timestamps from the $i^{\mbox{th}}$ master-slave communication path can be arranged in vector form as 
{\color{black}
\begin{align}\label{noattack_time_vecdata}
\bm{t}_{2i} = (\bm{t}_{1i} + d_i\mathds{1}_P + \bm{w}_{1i})\phi + \delta\mathds{1}_P & ; & \bm{t}_{3i} =  (\bm{t}_{4i} - d_i\mathds{1}_P - \bm{w}_{2i})\phi + \delta\mathds{1}_P.
\end{align}}
Similarly, when $\eta_i = 1$, the received timestamps from the $i^{\mbox{th}}$ master-slave communication path can be arranged in vector form as
{\color{black}
\begin{align}\label{attack_time_vecdata}
\bm{t}_{2i} = (\bm{t}_{1i} + (d_i + \tau_i)\mathds{1}_P + \bm{w}_{1i})\phi + \delta\mathds{1}_P & ; & \bm{t}_{3i} =  (\bm{t}_{4i} - d_i\mathds{1}_P - \bm{w}_{2i})\phi + \delta\mathds{1}_P.
\end{align}}
The complete set of received timestamps is denoted by $\bm{t} = [\bm{t}_{11}, \bm{t}_{12}, \cdots, \bm{t}_{1N}, \bm{t}_{21}, \bm{t}_{22}, \cdots$, $\bm{t}_{2N}, \bm{t}_{31}, \bm{t}_{32}, \cdots, \allowbreak \bm{t}_{3N},  \bm{t}_{41}, \bm{t}_{42}, \cdots, \bm{t}_{4N}]$. In our work, we seek estimators of $\delta$ and $\phi$ from the received timestamps $\bm{t}$, when $\bm{\eta}$ is unknown. We now state the assumptions made in our work.


\noindent
\textbf{{Assumption 1:}} \emph{{\color{black} Following \cite{Gaderer_2010, Mizrahi_2012, Mizrahi_2012_2, Shpiner_2013}, we assume the availability of $N$ master-slave communication paths and further assume that fewer than half of the $N$ master-slave communication paths are asymmetric, i.e., $||\bm{\eta}||_1 < N/2$. The latter assumption of $||\bm{\eta}||_1 < N/2$ is sufficient to identify the symmetric paths after clustering. Having some protected paths would be an alternative.}}

\noindent
\textbf{{Assumption 2:}} \emph{\color{black} All the queuing delays are strictly positive random variables and have finite support. Following \cite{Noh_2007, Chaudhari_2008, Chaudhari_MVUE_Info, Li_Jeske_2009, Wu_2011, PTP_LP, Anand_2015, Anand_2015_Lest, AK_Blum_offset, Karthik_Skew_Offset}, we assume that the random queuing delays $\{w_{kij}\}^P_{j=1}$ are independent and identically distributed in our work. The pdf of the random variables are denoted by $f_{ki}(.)$ for $i = 1, 2, \cdots , N$, $k = 1, 2$. Usually, the two-way timing message exchanges are sufficiently spaced apart in time to ensure that the random queuing delays are independent. Further the background traffic patterns in networks remain constant over several minutes. Hence, the assumption that all queuing delays in a particular master-slave communication path share a common pdf is fairly realistic. } 

\noindent
\textbf{{Assumption 3:}} \emph{\color{black} Following \cite{Noh_2007, Chaudhari_2008, Chaudhari_MVUE_Info, Wu_2011}, we assume the unknown deterministic path delays $\{d_i\}_{i=1}^N$, unknown biases $\{\tau_i\}_{i=1}^N$, clock skew $\phi$ and the clock offset $\delta$ are constant over $P$ two-way message exchanges for each master-slave communication path. } 

\noindent
\textbf{{Assumption 4:}} \emph{As very small $\tau_i$ will have little impact, we officially define a master-slave communication path as having an unknown asymmetry ($\eta_i = 1$) {\color{black} when} $|\tau_i| \ge d_{\tau}$, where $d_{\tau}$ can be chosen such that $|\tau_i| < d_{\tau}$ causes little impact.}

\section{Performance Lower Bounds For a Robust Clock Skew and Offset Estimation Scheme}\label{Sec3}
In this section, we develop useful performance lower bounds that help in evaluating the performance of the proposed clock skew and offset estimation schemes that are robust to unknown path asymmetries. We assume $\bm{\eta}$ is known and further assume complete knowledge of $f_{ki}(.)$ for $i = 1, 2, \cdots, N$ and $k = 1, 2$. We use invariant decision theory (see chapter $6$ of \cite{Berger}) to develop the optimum approach for fusing information from the $N$ master-slave communication paths. {\color{black} For ease of notation, we assume the first $K (< N/2)$ master-slave communication paths are asymmetric and the remaining $(N - K)$ master-slave communication paths are symmetric.} {\color{black} Under these assumptions with (\ref{noattack_time_vecdata}) and (\ref{attack_time_vecdata}), we obtain}
\begin{eqnarray}\label{vec_y_attack}
\bm{y}_i & = & (\bm{h}d_i + \bm{g}\tau_i + \bm{v}_i)\phi + \delta\mathds{1}_{2P},
\end{eqnarray}
for $i = 1, 2, \cdots, K$ and
\begin{eqnarray}\label{vec_y_noattack}
\bm{y}_i & = & (\bm{h}d_i + \bm{v}_i)\phi + \delta\mathds{1}_{2P},
\end{eqnarray}
for $i = K+1, \cdots, N$. In (\ref{vec_y_attack}) and (\ref{vec_y_noattack}), $\bm{y}_i = [\bm{t}_{2i}^T, \bm{t}_{3i}^T]^T$, $\bm{h} = [\mathds{1}_P^T, -\mathds{1}_P^T]^T$, $\bm{g} = [\mathds{1}_P^T, \bm{0}_P^T]^T$ and $\bm{v}_i = [\bm{v}_{1i}^T, \bm{v}_{2i}^T]^T$ with $\bm{v}_{1i} = (\bm{t}_{1i} + \bm{w}_{1i})$ and $\bm{v}_{2i} = (\bm{t}_{4i} - \bm{w}_{2i})$. The complete set of observations from the $N$ master-slave communication paths can be represented in vector form as
\begin{eqnarray}\label{MainModel}
\bm{y} & = & (\bm{H}\bm{\gamma} + \bm{v} )\phi + \delta\mathds{1}_{2NP},
\end{eqnarray}
where $\bm{y} = [\bm{y}_1^T, \bm{y}_2^T, \cdots, \bm{y}_N^T]$, $\bm{v} = [\bm{v}_1^T, \bm{v}_2^T, \cdots, \bm{v}_N^T]$ and $\bm{\gamma} = [\bm{d}, \bm{\tau}]$ with $\bm{d} = [d_1, d_2, \cdots, d_N]$ and $\bm{\tau} = [\tau_1, \tau_2, \cdots, \tau_K]$ and $\bm{H} = \begin{bmatrix}
\bm{h} \otimes \bm{I}_N, \bm{g} \otimes \bm{I}_K
\end{bmatrix}$. Let $\bm{\theta} = [\phi, \delta, d_1, d_2, \cdots, d_N, \tau_1, \cdots, \tau_K]$ denote the vector of unknown parameters. The parameter space of $\bm{\theta}$, denoted by $\bm{\Theta}$, is given by $\bm{\Theta} = \{(\phi, \delta, \bm{d}, \bm{\tau}): \phi \in \mathbb{R}^+, {\delta} \in \mathbb{R}, \bm{d} \in \mathbb{R}^N, \bm{\tau} \in \mathbb{R}^K  \}$. From (\ref{MainModel}), the conditional pdf of $\bm{y}$ is given by
\begin{align}
f(\bm{y}|\bm{\theta})  = & \frac{1}{\phi^{2NP}} f_{\bm{v}}\left( \frac{\bm{y} - \delta  \mathds{1}_{2NP}}{\phi}  - \bm{H}\bm{\gamma} \right), \\
 = & \frac{1}{\phi^{2NP}} \prod_{i=1}^{K} f_{\bm{v}_i}\left( \frac{\bm{y}_i - \delta  \mathds{1}_{2P}}{\phi}  - d_i \bm{h} - \tau_i \bm{g} \right)  \prod_{i=K+1}^{N} f_{\bm{v}_i}\left( \frac{\bm{y}_i - \delta  \mathds{1}_{2P}}{\phi}  - d_i \bm{h} \right), 
\end{align}
where $f_{\bm{v}_i}(\bm{v}_i) = \prod_{j=1}^{P} f_{1i} \left( {v}_{1ij} - {t}_{1ij} \right)f_{2i} \left( {t}_{4ij} - {v}_{2ij} \right)$ for $i = 1,2, \cdots, N$. Let $\mathcal{F}_{M}$ denote the class of all pdfs $f(\bm{y}|\bm{\theta})$ for $\bm{\theta} \in \bm{\Theta}$. The class of such pdfs is invariant under the group of transformations $\mathcal{G}_{M}$ on the observations $\bm{y}$, on $\mathbb{R}^{2NM}$, defined as
\begin{align}\label{M_model_group}
\mathcal{G}_{M}  & =  \{ g_{a, \bm{b}, c}(\bm{y}) : g_{a, \bm{b}, c}(\bm{y})  = (\bm{y} + \bm{H}\bm{b})a + c \mathds{1}_{2NP}, \forall (a, \bm{b}, c) \in \mathbb{R}^+ \times \mathbb{R}^{N+K} \times \mathbb{R}  \}, 
\end{align} 
where $\bm{y} \in \mathbb{R}^{2NM}$, since $\bm{y}_g = g_{a, \bm{b}, {c}}(\bm{y})$ has a pdf given by $\frac{1}{(a\phi)^{2NM}}$  $f_{\bm{v}}\left( \frac{\bm{y}_g - ((a{\delta} + {c}) \mathds{1}_{2NP})}{a\phi} \allowbreak  - \bm{H}\left( \bm{\gamma} + \frac{\bm{b}}{\phi} \right)\right)$. The corresponding group of induced transformations on $\bm{\Theta}$, denoted by $\bar{\mathcal{G}}_{M}$, is given by 
\begin{align}\label{M_model_param_group}
\bar{\mathcal{G}}_{M} & = \{ \bar{g}_{a,\bm{b},{c}}((\phi, \bm{\gamma}, {\delta})) : \bar{g}_{a, \bm{b}, {c}}((\phi, \bm{\gamma}, {\delta})) = (a\phi,  (\bm{\gamma} + \bm{b}/\phi), (a{\delta} + {c})),  \forall (a, \bm{b}, c) \in \mathbb{R}^+ \times \mathbb{R}^{N+K} \times \mathbb{R}  \},
\end{align}
where $\phi \in \mathbb{R}^+, \bm{\gamma} \in \mathbb{R}^{N+K}$ and ${\delta} \in \mathbb{R}$.

Let $\hat{\delta}_I$ and $\hat{\phi}_I$ denote estimators of $\delta$ and $\phi$, respectively and let $\hat{\delta}_I(\bm{y})$ and $\hat{\phi}_I(\bm{y})$ denote the estimates obtained from the received data $\bm{y}$ characterized by the pdf $f(\bm{y}|\bm{\theta}) = \frac{1}{\phi^{2NP}} \allowbreak f_{\bm{v}}\left( \frac{\bm{y} - ({\delta} \mathds{1}_{2NP})}{\phi} - \bm{H}\bm{\gamma}\right)$. From (\ref{M_model_param_group}), the estimators $\hat{\phi}_{I}(\bm{y})$ and $\hat{\delta}_{I}(\bm{y})$ are invariant under $\mathcal{G}_{M}$ from (\ref{M_model_group}), if for all $(a, \bm{b}, {c}) \in \mathbb{R}^+ \times \mathbb{R}^{N+K} \times \mathbb{R}$,
\begin{align}
\hat{\delta}_{I}(g_{a, \bm{b}, c}(\bm{y})) = \hat{\delta}_{I}(a(\bm{y} + \bm{Hb}) + {c}\mathds{1}_{2NP}) & =  a\hat{\delta}_{I}(\bm{y}) + c, \label{delta_invariant}\\
\hat{\phi}_{I}(g_{a, \bm{b}, {c}}(\bm{y})) = \hat{\phi}_{I}(a(\bm{y} + \bm{Hb}) + {c} \mathds{1}_{2NP}) & =  a\hat{\phi}_{I}(\bm{y}). \label{phi_invariant}
\end{align}
In this paper, we consider the skew-normalized squared error loss functions for $\delta$ and $\phi$ defined by $\frac{(\hat{\delta}_{I}(\bm{y}) - \delta)^2}{\phi^2}$ and $\frac{(\hat{\phi}_{I}(\bm{y}) - \phi)^2}{\phi^2}$, respectively. The corresponding conditional risk for $\hat{\delta}_I$ and $\hat{\phi}_I$ under the skew normalized square error loss functions are the skew-normalized mean square estimation errors, defined by
\begin{align}
\mathcal{R}(\hat{\delta}_I, \bm{\theta}) =  \frac{1}{\phi^2}\int_{\mathbb{R}^{2NM}} (\hat{\delta}_I(\bm{y}) - \delta)^2 f(\bm{y}| \bm{\theta}) d\bm{y} & \mbox{;} & \mathcal{R}(\hat{\phi}_I, \bm{\theta}) =  \frac{1}{\phi^2}\int_{\mathbb{R}^{2NM}} (\hat{\phi}_I(\bm{y}) - \phi)^2 f(\bm{y}| \bm{\theta}) d\bm{y},\label{NRMSE_offset}
\end{align}
respectively. The skew-normalized loss functions for $\delta$ and $\phi$ are invariant under $\mathcal{G}_{M}$ from (\ref{M_model_group}), since
\begin{align}
\frac{(\hat{\delta}_{I}(\bm{y}) - \delta)^2}{\phi^2} =  \frac{\left(\hat{\delta}_{I}(g_{a, \bm{b}, {c}}(\bm{y})) - (a\delta + c)\right)^2}{a^2\phi^2} \mbox{ and } \frac{(\hat{\phi}_{I}(\bm{y}) - \phi)^2}{\phi^2} =  \frac{\left(\hat{\phi}_{I}(g_{a, \bm{b}, {c}}(\bm{y})) - a\phi\right)^2}{a^2\phi^2}, 
\end{align}
for all $g_{a, \bm{b}, {c}} \in \mathcal{G}_{M}$. We now present the optimum invariant (or minimum conditional risk) estimators of $\delta$ and $\phi$.

\begin{proposition}(Genie Bound)\label{Minimax_optimum_estimator_phase_freq}
	Assuming knowledge of the paths having an unknown asymmetry and complete knowledge of $f_{1i}(.)$ and $f_{2i}(.)$ for $i = 1, 2, \cdots, N$, the optimum invariant estimators for $\delta$ and $\phi$, denoted by $\hat{\delta}_{opt}$ and $\hat{\phi}_{opt}$, respectively, are given by 
	\begin{align}
\hat{\delta}_{opt}(\bm{y})  & = \frac{ \int_{\mathbb{R}^+}\int_{\mathbb{R}^{N+K+1}} \frac{\delta \Gamma_1(\phi, \delta, \bm{d}, \bm{\tau}, \bm{y}) \Gamma_0(\phi, \delta, \bm{d}, \bm{y})}{\phi^{2NP - N - K + 3}} d\bm{\tau} d(\bm{d}) d\delta d\phi }{ \int_{\mathbb{R}^+}\int_{\mathbb{R}^{N+K+1} } \frac{\Gamma_1(\phi, \delta, \bm{d}, \bm{\tau}, \bm{y}) \Gamma_0(\phi, \delta, \bm{d}, \bm{y})}{\phi^{2NP - N - K + 3}}   d\bm{\tau} d(\bm{d}) d\delta d\phi}, \label{Minimax_offset}
	\end{align}
	and
	\begin{align}
\hat{\phi}_{opt}(\bm{y})  & = \frac{ \int_{\mathbb{R}^+}\int_{\mathbb{R}^{N+K+1}} \frac{\Gamma_1(\phi, \delta, \bm{d}, \bm{\tau}, \bm{y}) \Gamma_0(\phi, \delta, \bm{d}, \bm{y})}{\phi^{2NP - N - K + 2}}   d\bm{\tau} d(\bm{d}) d\delta d\phi }{ \int_{\mathbb{R}^+}\int_{\mathbb{R}^{N+K+1}} \frac{\Gamma_1(\phi, \delta, \bm{d}, \bm{\tau}, \bm{y}) \Gamma_0(\phi, \delta, \bm{d}, \bm{y})}{\phi^{2NP - N - K + 3}}   d\bm{\tau} d(\bm{d}) d\delta d\phi }, \label{Minimax_skew}
	\end{align}
	respectively, where we have $\Gamma_1(\phi, \delta, \bm{d}, \bm{\tau}, \bm{y}) = \prod_{i=1}^{K} f_{\bm{v}_i}\left( \frac{\bm{y}_i - \delta  \bm{1}_{2M}}{\phi}  - d_i \bm{h} - \tau_i \bm{g} \right)$ and $\Gamma_0(\phi, \delta, \bm{d}, \bm{y}) = \allowbreak \prod_{i=K+1}^{N} \allowbreak f_{\bm{v}_i}\left( \frac{\bm{y}_i - \delta  \bm{1}_{2M}}{\phi}  - d_i \bm{h} \right)$. Following a proof similar to that given in \cite{Karthik_Skew_Offset}, we can show the estimators $\hat{\delta}_{opt}$ and $\hat{\phi}_{opt}$ are minimax optimum, i.e., {\color{black} they minimize the maximum of the} skew-normalized mean square estimation error (NMSE) over all parameter values. As these optimum estimators achieve the smallest NMSE among the class of invariant estimators and are minimax optimum, the performance of these estimators give us useful fundamental lower bounds on the skew-normalized mean square estimation error for a clock skew and offset estimation scheme. {\color{black} We refer to  the clock skew and offset estimators presented in (\ref{Minimax_offset}) and (\ref{Minimax_skew}) as genie optimum estimators\footnote{These estimators are called genie optimum estimators since they assume the availability of information that is not usually available and are optimum in terms of minimizing the NMSE among the class of invariant estimators.}.}
\end{proposition}

{\color{black} In certain scenarios where we have analytical expressions for $f_{1i}(.)$ and $f_{2i}(.)$ for $i = 1, 2, \cdots, N$, it might be possible to further simplify the integrals in (\ref{Minimax_offset}) and (\ref{Minimax_skew}). However, for the general case of arbitrary queuing delay pdfs, these integrals are computed by approximating them with Riemann summations\footnote{In our work, we consider the LTE backhaul network scenario. The empirical pdfs of the random queuing delays for this scenario was derived in \cite{Anand_2015}. We use the empirical pdfs from \cite{Anand_2015} to calculate (\ref{Minimax_offset}) and (\ref{Minimax_skew}).}. The width of the Riemann summation bins is set to small values to ensure that the additional error introduced due to the Riemann sum approximation is insignificant relative to the estimation error. }

\begin{remark}
	{\color{black} Proposition 1 provides us with mathematical expressions for the optimum clock skew and offset estimators for IEEE 1588 under the assumption that we have complete knowledge of $f_{1i}(.)$ and $f_{2i}(.)$ for $i = 1, 2, \cdots, N$ and prior knowledge of the master-slave communication paths having unknown deterministic path asymmetries. The skew normalized mean square estimation error (NMSE) performance of the optimum estimators described in Proposition 1 cannot be achieved unless we have information that is usually not available (prior information regarding the master-slave communication paths having unknown deterministic path delay asymmetries). Hence, the NMSE performance of the optimum clock skew and offset estimator described in Proposition 1 can be viewed as a performance lower bound.  The NMSE performance of the optimum clock skew and offset estimator described in Proposition 1 can be viewed as a performance lower bound as it gives us {a lower bound on} the NMSE for invariant clock skew and offset estimation schemes in the presence of possible unknown deterministic path delay asymmetries. }
\end{remark}

\section{Robust Clock Skew and Offset Estimation Scheme}\label{Sec4}
In this section, we present our robust scheme for jointly estimating the clock skew and offset in the presence of master-slave communication paths with possible unknown asymmetries. When developing the performance bounds {\color{black} in Proposition \ref{Minimax_optimum_estimator_phase_freq}}, we had assumed prior information regarding the paths having an unknown asymmetry as well as the complete knowledge of the distribution of the queuing delays. However in practice, we generally do not have  information regarding the asymmetric master-slave communication paths. Hence, {\color{black} we attempt} to identify these paths when developing a robust clock skew and offset scheme. Further in some scenarios, we might not have the complete information regarding the pdf of the random queuing delays, $f_{1i}(.)$ and $f_{2i}(.)$ for $i = 1, 2, \cdots, N$. In this paper, we use the popular Gaussian mixture model (GMM) \cite{Blum_spatial, Kostantinos} for approximating the pdf of the random queuing delays as\footnote{\color{black} The GMM is known to be a universal approximator in the sense discussed in \cite{Kariya}.}
\begin{align}
f_{1i}(w)   =   \sum_{k=1}^{M_i} \alpha_{ik} \mathcal{P}_{\mu_{1ik}, \sigma_{1ik}}\left(w  \right)& ;& f_{2i}(w)   = \sum_{l=1}^{L_i} \beta_{il} \mathcal{P}_{\mu_{2il}, \sigma_{2il}}\left(w  \right),  \label{fwd_mixture} 
\end{align}
for $i = 1, 2, \cdots, N$. In (\ref{fwd_mixture}), $\{\alpha_{ik}\}_{k=1}^{M_i}$ and $\{\beta_{il}\}_{l=1}^{L_i}$ denote the unknown mixing coefficients in the forward and reverse path at the $i^{\mbox{th}}$ master-slave communication path with $M_i$ and $L_i$ denoting the number of {\color{black} assumed} mixture components in the forward and reverse path, respectively. Also, we have $\alpha_{ik} \in [0, 1]$ and $\beta_{il} \in [0, 1]$ with the constraints $\sum_{k=1}^{M_i} \alpha_{ik} = 1$ and $\sum_{l=1}^{L_i} \beta_{il} = 1$. Further, $\mathcal{P}_{\mu, \sigma}(.)$ denotes a normal distribution with mean $\mu$ and standard deviation $\sigma$. The variables $\{\mu_{1ik}, \sigma_{1ik}\}$ denote the mean and standard deviation of the $k^{\mbox{th}}$ component in the mixture {\color{black} models} in the $i^{\mbox{th}}$ forward path and the variables $\{\mu_{2il}, \sigma_{2il}\}$ denote the mean and standard deviation of the $l^{\mbox{th}}$ component in the mixture {\color{black} models} in the $i^{\mbox{th}}$ reverse path. 
{\color{black} 
A set of samples $\tilde{\bm{w}}_{k} = [\tilde{\bm{w}}_{k1}, \tilde{\bm{w}}_{k2}, \cdots, \tilde{\bm{w}}_{kN}]$ for $k = 1, 2$, where $\tilde{\bm{w}}_{ki} = [\tilde{{w}}_{ki1}, \tilde{{w}}_{ki2}, \cdots, \tilde{{w}}_{kiP_t}]$ for $i = 1, 2, \cdots, N$ are obtained from the previous block of $P_t$ two-way message exchanges which we call the previous window as we describe next\footnote{The samples $\tilde{\bm{w}}_{k}$ cannot be obtained in certain scenarios such as the initial startup or when the network routes have changed drastically from the previous window to the current window which will be reported by the router. There are startup routines that are currently used in practice. 
They typically employ a large number of two-way message exchanges and possibly several iterative improvement stages.}.  The full impact of using these samples is characterized in Section V, where it is shown that in the cases studied, they provide positive impact. Let the received timestamps corresponding to the previous block be denoted by $\tilde{\bm{t}}_{ki} = [\tilde{t}_{ki1}, \tilde{t}_{ki2}, \cdots,$ $\tilde{t}_{kiP_t}]$ for $k = 1, 2, 3, 4$ and $i = 1, 2, \cdots, N$ and the previous estimates of the $\delta$, $\phi$, $d_i^{ms}$ and $d_i^{sm}$ for $i = 1, 2, \cdots, N$ be denoted by $\tilde{\delta}^{old}$, $\tilde{\phi}^{old}$, $\tilde{d}_{i}^{ms}$ and $\tilde{d}_{i}^{sm}$ for $i = 1, 2, \cdots, N$, respectively. From (1), we then obtain $\tilde{\bm{w}}_{k}$ for $k = 1, 2$ as
\begin{align}
    \tilde{\bm{w}}_{1i}  = \frac{\tilde{\bm{t}}_{2i} - \tilde{\delta}^{old}\mathds{1}_{P_t}}{\tilde{\phi}^{old}} - \tilde{\bm{t}}_{1i} - \tilde{d}_{i}^{ms}\mathds{1}_{P_t}; &
    \qquad \tilde{\bm{w}}_{2i}  = \frac{\tilde{\delta}^{old}\mathds{1}_{P_t} - \tilde{\bm{t}}_{3i}}{\tilde{\phi}^{old}} + \tilde{\bm{t}}_{4i} - \tilde{d}_{i}^{sm}\mathds{1}_{P_t}    
\end{align}
for $i = 1, 2, \cdots, N.$ }

Let $\bm{\Omega}$ denote the vector of unknown parameters defined as 
$\bm{\Omega} = [\bm{\Psi}, \bm{\eta}, \bm{\alpha}_1, \cdots, \bm{\alpha}_N, \bm{\beta}_1, \cdots, \bm{\beta}_N, \bm{\mu}_{11}, \cdots$  $\bm{\mu}_{1N},  \bm{\sigma}_{11}, \cdots, \bm{\sigma}_{1N}, \bm{\mu}_{21}, \cdots, \bm{\mu}_{2N}, \bm{\sigma}_{21}, \cdots, \bm{\sigma}_{2N}]$
where we have $\bm{\Psi} = [\phi, \delta, d_1, \cdots, d_N, \tau_1, \cdots, \tau_N]$, $\bm{\eta} = [\eta_1, \eta_2, \cdots, \eta_N]$, $\bm{\alpha}_i = [\alpha_{i1}, \cdots, \alpha_{iM_i}]$ for $i = 1, 2, \cdots, N$, $\bm{\beta}_i = [\beta_{i1}, \cdots, \beta_{iL_i}]$ for $i = 1, 2, \cdots, N$, $\bm{\mu}_{1i} = [\mu_{11}, \cdots, \mu_{1M_i}]$ for $i = 1, 2, \cdots, N$, $\bm{\mu}_{2i} = [\mu_{21}, \cdots, \mu_{2L_i}]$ for $i = 1, 2, \cdots, N$, $\bm{\sigma}_{1i} = [\sigma_{1i1}, \cdots, \sigma_{1iM_i}]$ for $i = 1, 2, \cdots, N$ and $\bm{\sigma}_{2i} = [\sigma_{2i1}, \cdots, \sigma_{2iL_i}]$ for $i = 1, \cdots, N$.  {\color{black} Given $\bm{\Omega}$,} the log-likelihood function of the observed data $\bm{t}$, $\tilde{\bm{w}}_1$ and $\tilde{\bm{w}}_2$, denoted by $\mathcal{L}(\bm{\Omega}|\bm{t}, \tilde{\bm{w}}_1, \tilde{\bm{w}}_2)$, is defined as 
{\small
\begin{align}
& \sum_{i=1}^N \sum_{j=1}^P \Bigg[ \ln \left( \eta_i \Bigg( \sum_{k=1}^{M_i} \alpha_{ik} \mathcal{P}_{\mu_{1ik}, \sigma_{1ik}} \left( \frac{t_{2ij} - \delta }{\phi}  - d_i - \tau_i \right. \right. \left. - t_{1ij} \right) \Bigg)\left. \Bigg(\sum_{l=1}^{L_i} \beta_{il} \mathcal{P}_{\mu_{2il}, \sigma_{2il}} \left( t_{4ij} - d_i + \frac{\delta - t_{3ij} }{\phi} \right)\right) \Bigg)  \nonumber \\
& +  (1 - \eta_i) \left(\sum_{k=1}^{M_i} \alpha_{ik} \mathcal{P}_{\mu_{1ik}, \sigma_{1ik}} \left( \frac{t_{2ij} - \delta }{\phi}  - d_i - t_{1ij} \right)\right.\Bigg)  \left. \Bigg( \sum_{l=1}^{L_i} \beta_{il} \mathcal{P}_{\mu_{2il}, \sigma_{2il}} \left( t_{4ij} - d_i + \frac{\delta - t_{3ij} }{\phi} \right)\right) \Bigg) \Bigg]   \nonumber \\
& + \sum_{i=1}^N \sum_{j=1}^{P_t}  \ln\left( \sum_{k=1}^{M_i} \alpha_{ik}  \mathcal{P}_{\mu_{1ik}, \sigma_{1ik}} (\tilde{w}_{1ij}) \right)  + \sum_{i=1}^N \sum_{j=1}^{P_t}  \ln\left( \sum_{l=1}^{L_i} \beta_{il}  \mathcal{P}_{\mu_{2il}, \sigma_{2il}} (\tilde{w}_{2ij}) \right) - 2NM\ln\phi. \label{binary_completeLL}
\end{align}
}
The maximum likelihood method is widely used and has many attractive features including consistency and asymptotic unbiasedness.  {\color{black} Under assumptions $1-3$,} the maximum likelihood estimate (MLE) of $\bm{\Omega}$, denoted by $\hat{\bm{\Omega}}_{mle}$, is obtained by solving the following constrained optimization problem.
\begin{subequations}\label{main_equation}
	\begin{align}
	\hat{\bm{\Omega}}_{mle} & = \operatorname*{arg\,max}_{\bm{\Omega}}  \mathcal{L}(\bm{\Omega}|\bm{t}, \tilde{\bm{w}}_{1}, \tilde{\bm{w}}_{2}) \tag{\ref{main_equation}} \\
	\mbox{ such that }\eta_i & \in \{0, 1\} \mbox{ for } i = 1, 2, \cdots, N, \label{Con1} \\
	\alpha_{ik} & \in [0, 1] \mbox{ for } i = 1, 2, \cdots, N,   k = 1, 2, \cdots, M \mbox{, } \sum_{k=1}^{M_i} \alpha_{ik} = 1, \label{Con2} \\
	\beta_{il} & \in [0, 1] \mbox{ for } i = 1, 2, \cdots, N  \mbox{, }  l = 1, 2, \cdots, M \mbox{, } \sum_{l=1}^{L_i} \beta_{il} = 1, \label{Con3} \\
	|\tau_i| & \ge d_{\tau} \mbox{ when } \eta_i = 1, \label{Con5} \\
	\sum_{i=1}^N \eta_i & \le N/2. \label{Con6}
	\end{align}
\end{subequations}
\noindent
The mixed integer nonlinear programming problem presented in (\ref{main_equation}) is computationally intensive to solve for large values of $N$ as we would have {\color{black} to generally} search across $2^N$ possibilities of $\bm{\eta}$. In this paper, we use the idea discussed in \cite{Zhang_EM_CVTR} to solve a relaxed version of (\ref{main_equation}) and to obtain a robust estimate of the clock skew and offset\footnote{In \cite{Zhang_EM_CVTR}, the authors address the problem of data fusion in the presence of attacks in wireless sensor networks for IoT applications. They do not look at the problem of clock skew and offset estimation for IEEE 1588 in the presence of unknown deterministic path asymmetries. }.

\subsection{Binary Variable Relaxation and EM algorithm}
As the constraints in (\ref{Con1}) correspond to binary variables, we relax the problem and introduce real variables with constraints defined as $\pi_i = \mbox{Pr}(\eta_i = 1) \in (0, 1)$ for $i = 1, 2, \cdots, N$. Let $\bm{\Omega}_{\pi} = [\bm{\Psi}, \bm{\pi}, \bm{\alpha}_1, \cdots, \bm{\alpha}_N,\allowbreak \bm{\beta}_1, \cdots, \bm{\beta}_N, \bm{\mu}_{11},  \cdots, \bm{\mu}_{1N}, \bm{\sigma}_{11}, \cdots, \bm{\sigma}_{1N}, \allowbreak \bm{\mu}_{21}, \cdots, \bm{\mu}_{2N}, \bm{\sigma}_{21}, \cdots, \bm{\sigma}_{2N}]$, where we have $\bm{\pi} = [\pi_1, \pi_2, \cdots, \pi_N]$. Replacing the binary variables with the corresponding real variables and dropping the constraints in (\ref{Con5}) and (\ref{Con6}), we can rewrite the optimization problem in (\ref{main_equation}) as
\begin{subequations}\label{main_equation_EM}
	\begin{align}
	\hat{\bm{\Omega}}_{\pi, mle} & = \operatorname*{arg\,max}_{\bm{\Omega}_{\pi}} \mathcal{L}_{EM}({\bm{\Omega}}_{\pi}|\bm{t}, \tilde{\bm{w}}_1, \tilde{\bm{w}}_2) \tag{\ref{main_equation_EM}} \\
	\mbox{ such that } \pi_i & \in (0, 1) \mbox{ for } i = 1, 2, \cdots, N, \label{Con1_EM} \\
	\alpha_{ik} & \in (0, 1) \mbox{ with } \sum_{k=1}^{M_i} \alpha_{ik} = 1 \mbox{ for } i = 1, 2, \cdots, N, \label{Con2_EM} \\
	\beta_{il} & \in (0, 1) \mbox{ with } \sum_{l=1}^{L_i} \beta_{il} = 1 \mbox{ for } i = 1, 2, \cdots, N,  \label{Con3_EM}
	\end{align}
\end{subequations}
where $\hat{\bm{\Omega}}_{\pi, mle}$ denotes the MLE of $\bm{\Omega}_{\pi}$ and $\mathcal{L}_{EM}({\bm{\Omega}}_{\pi}|\bm{t}, \tilde{\bm{w}}_1, \tilde{\bm{w}}_2)$, referred to as the incomplete log-likelihood is defined as
{\small
\begin{align}\label{Incomplete_logLL}
& \sum_{i=1}^N \sum_{j=1}^P \ln \left[\pi_i\left( \Bigg(\sum_{k=1}^{M_i} \alpha_{ik} \mathcal{P}_{\mu_{1ik}, \sigma_{1ik}}  \left( \frac{t_{2ij} - \delta }{\phi}  - d_i  - \tau_i - t_{1ij} \right) \right. \right.\Bigg)\left. \Bigg( \sum_{l=1}^{L_i} \beta_{il} \mathcal{P}_{\mu_{2il}, \sigma_{2il}}  \left( t_{4ij} - d_i + \frac{\delta - t_{3ij} }{\phi} \right)\right)\Bigg) \nonumber \\
& + (1 - \pi_i)\Bigg(\left(\sum_{k=1}^{M_i} \alpha_{ik} \mathcal{P}_{\mu_{1ik}, \sigma_{1ik}} \left( \frac{t_{2ij} - \delta }{\phi}  - d_i - t_{1ij} \right)\right.\Bigg) \left. \left. \Bigg(\sum_{l=1}^{L_i} \beta_{il} \mathcal{P}_{\mu_{2il}, \sigma_{2il}} \left( t_{4ij} - d_i +  \frac{\delta - t_{3ij} }{\phi} \right)\right)\Bigg)\right]   \nonumber \\
& + \sum_{i=1}^N \sum_{j=1}^{P_t}  \ln\left[ \sum_{k=1}^{M_i} \alpha_{ik}  \mathcal{P}_{\mu_{1ik}, \sigma_{1ik}} (\tilde{w}_{1ij}) \right]  + \sum_{i=1}^N \sum_{j=1}^{P_t} \ln \left[ \sum_{l=1}^{L_i} \beta_{il}  \mathcal{P}_{\mu_{2ik}, \sigma_{2ik}} (\tilde{w}_{2ij}) \right] - 2NM\ln\phi.
\end{align}
}
{\color{black}
	
	 The iterative algorithm for solving (\ref{main_equation_EM}) {\color{black} is next} enumerated in steps $1)-16)$. The SAGE algorithm proposed in \cite{Fessler} is used to derive steps $1)-16)$, and the details are presented in Appendix \ref{App_sec2}. The algorithm begins with the current estimates $\hat{\bm{\Omega}}_{\pi}^{'}$ of ${\bm{\Omega}}_{\pi}$ and produces updated estimates of ${\bm{\Omega}}_{\pi}$ as follows:

\begin{enumerate}
	\item {\color{black} In this step, we calculate the variables $D_{ij}$, $\chi_{ijkl}^{(1)}$ and $\chi_{ijkl}^{(0)}$ based on the current parameter estimates, $\hat{\bm{\Omega}}_{\pi}^{'}$, and the observed timestamps. These variables are necessary for calculating the updated estimates of the parameters in ${\bm{\Omega}}_{\pi}$.} Define $D_{ij}$ as
	\begin{align}
	D_{ij} =  & \sum_{k_c=1}^{M_i} \sum_{l_c=1}^{L_i} \Bigg[\hat{\pi}_i^{'}\hat{\alpha}_{ik_c}^{'} \hat{\beta}_{il_c}^{'}  \mathcal{P}_{\mu^{'}_{2il_c}, \sigma^{'}_{2il_c}} \left(t_{4ij} - \hat{d}_i^{'} + \frac{\hat{\delta}^{'} - t_{3ij} }{\hat{\phi}^{'}} \right)  \mathcal{P}_{\mu^{'}_{1ik_c}, \sigma^{'}_{1ik_c}} \left(\frac{t_{2ij} - \hat{\delta}^{'} }{\hat{\phi}^{'}}  - \hat{d}_i^{'} - \hat{\tau}_i^{'} - t_{1ij}\right) \nonumber \\
	& + (1 - \hat{\pi}_i^{'})\hat{\alpha}_{ik_c}^{'} \hat{\beta}_{il_c}^{'} \mathcal{P}_{\mu^{'}_{2il_c}, \sigma^{'}_{2il_c}} \left(t_{4ij} - \hat{d}_i^{'} + \frac{\hat{\delta}^{'} - t_{3ij} }{\hat{\phi}^{'}} \right)  \mathcal{P}_{\mu^{'}_{1ik_c}, \sigma^{'}_{1ik_c}} \left(\frac{t_{2ij} - \hat{\delta}^{'} }{\hat{\phi}^{'}}  - \hat{d}_i^{'} - t_{1ij}\right)\Bigg]. \nonumber
	\end{align}
	for $i = 1, 2, \cdots, N$ and $j = 1, 2, \cdots, P$. Then, compute 
	\begin{align}
	\chi_{ijkl}^{(1)} =  D_{ij}^{-1} \hat{\pi}_i^{'}\hat{\alpha}_{ik}^{'}\hat{\beta}_{il}^{'}\mathcal{P}_{\mu^{'}_{2il}, \sigma^{'}_{2il}} \left(t_{4ij} - \hat{d}_i^{'} + \frac{\hat{\delta}^{'} - t_{3ij} }{\hat{\phi}^{'}} \right) \mathcal{P}_{\mu^{'}_{1ik}, \sigma^{'}_{1ik}} \left(\frac{t_{2ij} - \hat{\delta}^{'} }{\hat{\phi}^{'}}  - \hat{d}_i^{'} - \hat{\tau}_i^{'} - t_{1ij}\right) \label{a1_ijkl_men1}
	\end{align}	
	and
	\begin{align}
	\chi_{ijkl}^{(0)} =  D_{ij}^{-1} (1 - \hat{\pi}_i^{'})\hat{\alpha}_{ik}^{'}\hat{\beta}_{il}^{'} 
	 \mathcal{P}_{\mu^{'}_{2il}, \sigma^{'}_{2il}} \left(t_{4ij} - \hat{d}_i^{'} + \frac{\hat{\delta}^{'} - t_{3ij} }{\hat{\phi}^{'}} \right) \mathcal{P}_{\mu^{'}_{1ik}, \sigma^{'}_{1ik}} \left(\frac{t_{2ij} - \hat{\delta}^{'} }{\hat{\phi}^{'}}  - \hat{d}_i^{'} - t_{1ij}\right) \label{a0_ijkl_men1}
	\end{align}	
	for $i = 1, 2, \cdots, N$, $j = 1, 2 \cdots, P$, $k = 1, 2, \cdots, M_i$ and $l = 1, 2, \cdots, L_i$.

	\item {\color{black} Similar to the first step, we calculate the variables $\tilde{D}_{ij}$ and $\tilde{a}_{ijkl}$ based on the current parameter estimates, $\hat{\bm{\Omega}}_{\pi}^{'}$, and the observed timestamps. These variables are necessary for calculating the updated estimates of parameters in ${\bm{\Omega}}_{\pi}$.} First, we calculate $\tilde{D}_{ij} = \sum_{k_c=1}^{M_i} \sum_{l_c=1}^{L_i} \hat{\alpha}_{ik_c}^{'} \hat{\beta}_{il_c}^{'}\mathcal{P}_{\mu^{'}_{2ik_c}, \sigma^{'}_{2ik_c}} \left(\tilde{w}_{1ij} \right) \allowbreak \mathcal{P}_{\mu^{'}_{2il_c}, \sigma^{'}_{2il_c}} \left(\tilde{w}_{2ij} \right)$ for $i = 1, 2, \cdots, N$ and $j = 1, 2, \cdots, P_t$. We then compute  	
	\begin{align}
	\tilde{a}_{ijkl} =  \tilde{D}_{ij}^{-1} \hat{\alpha}_{ik}^{'}\hat{\beta}_{il}^{'} \mathcal{P}_{\mu^{'}_{2ik}, \sigma^{'}_{2ik}} \left(\tilde{w}_{1ij} \right)  \mathcal{P}_{\mu^{'}_{2il}, \sigma^{'}_{2il}} \left(\tilde{w}_{2ij} \right) \label{a_noise_ijkl_men1}
	\end{align}	
	for $i = 1, 2, \cdots, N$, $j = 1, 2 \cdots, P_t$, $k = 1, 2, \cdots, M_i$ and $l = 1, 2, \cdots, L_i$. 

	\item {\color{black} In this step, we calculate the updated estimate of ${\pi}_i$, $\alpha_{ik}$ and $\beta_{il}$, denoted by ${\pi}_i^{'}$, $\alpha_{ik}^{'}$ and $\beta_{il}^{'}$, respectively, for $i = 1, 2, \cdots, N, k = 1, 2, \cdots, M_i$ and $l = 1, 2, \cdots, L_i$. We have}
	\begin{align}\label{pi_update_men1}
	\hat{\pi}_i = \frac{1}{P} \sum_{j=1}^P \sum_{k=1}^{M_i} \sum_{l=1}^{L_i} \chi_{ijkl}^{(1)},
	\end{align}	
	\begin{align}\label{alpha_update_men1}
	\hat{\alpha}_{ik} & = \frac{1}{(P + P_t)} \left[ \sum_{j=1}^P \sum_{l=1}^{L_i} \left(  \chi_{ijkl}^{(1)} + \chi_{ijkl}^{(0)} \right) \right.  + \left. \sum_{j=1}^{P_t} \sum_{l=1}^{L_i} \tilde{a}_{ikjl} \right],
	\end{align}
	\begin{align}\label{beta_update_men1}
	\hat{\beta}_{il} & = \frac{1}{(P + P_t)} \left[ \sum_{j=1}^P \sum_{k=1}^{M_i} \left(  \chi_{ijkl}^{(1)} + \chi_{ijkl}^{(0)} \right) \right. + \left. \sum_{j=1}^{P_t} \sum_{k=1}^{M_i} \tilde{a}_{ikjl} \right]. 
	\end{align}
	
	\item {\color{black} In this step, we update the current estimates of ${\pi}_i$, $\alpha_{ik}$ and $\beta_{il}$ with the estimates obtained from step 3 and we recompute the variables in steps 1 \& 2. } Set $\hat{\pi}_{i}^{'} = \hat{\pi}_{i}$,  $\hat{\alpha}_{ik}^{'} = \hat{\alpha}_{ik}$ and $\hat{\beta}_{il}^{'} = \hat{\beta}_{il}$ for $i = 1, 2, \cdots, N, k = 1, 2, \cdots, M_i$ and $l = 1, 2, \cdots, L_i$ {\color{black} and recompute $D_{ij}$ and $\tilde{D}_{ij}$. Then recompute $\chi_{ijkl}^{(1)}$, $\chi_{ijkl}^{(0)}$ and $\tilde{a}_{ijkl}$ using (\ref{a1_ijkl_men1}), (\ref{a0_ijkl_men1}) and (\ref{a_noise_ijkl_men1}), respectively.}	 
	
	\item {\color{black} In this step, we calculate the updated estimates of $\mu_{1ik}$ and ${\mu}_{2il}$, denoted by $\mu_{1ik}^{'}$ and ${\mu}_{2il}^{'}$, respectively, for $i = 1, 2, \cdots, N$, $k = 1, 2, \cdots, M_i$ and $l = 1, \cdots, L_i$.} Define  $D_{\mu_1, ik} = \sum_{j=1}^P \sum_{l=1}^{L_i} \left( \chi_{ijkl}^{(1)} + \chi_{ijkl}^{(0)}\right) + \sum_{j=1}^{P_t} \sum_{l=1}^{L_i} \tilde{a}_{ijkl}$ and $D_{\mu_2, il} = \sum_{j=1}^P \sum_{k=1}^{M_i} \left( \chi_{ijkl}^{(1)} + \chi_{ijkl}^{(0)}\right) + \sum_{j=1}^{P_t} \sum_{k=1}^{M_i} \tilde{a}_{ijkl}$ for $i = 1, 2, \cdots, N$, $k = 1, 2, \cdots, M_i$ and $l = 1, \cdots, L_i$. Then compute
	\begin{align}
	\hat{\mu}_{1ik} =  D_{\mu_1, ik}^{-1} \left[ \sum_{j=1}^P \sum_{l=1}^{L_i} \Bigg( \bigg(\frac{t_{2ij} - \hat{\delta}^{'}}{\hat{\phi}^{'}} - \hat{d}_i^{'}  -t_{1ij}\bigg)\right. \left( \chi_{ijkl}^{(1)} + \chi_{ijkl}^{(0)}\right) - \chi_{ijkl}^{(1)}\hat{\tau}_i^{'} \Bigg) \left. + \sum_{j=1}^{P_t} \sum_{l=1}^{L_i} \tilde{a}_{ijkl}\tilde{w}_{1ij}\right], \label{mu_1ik}
	\end{align}
and
	\begin{align}
	\hat{\mu}_{2il} = & D_{\mu_2, il}^{-1} \bigg[   \sum_{j=1}^P \sum_{k=1}^{M_i}  \left( \chi_{ijkl}^{(1)} + \chi_{ijkl}^{(0)}\right) \bigg(t_{4ij} - \hat{d}_i^{'}  + \frac{\hat{\delta}^{'} - t_{3ij}}{\hat{\phi}^{'}}\bigg)+ \sum_{j=1}^{P_t} \sum_{k=1}^{M_i} \tilde{a}_{ijkl}\tilde{w}_{2ij}\bigg] \label{mu_2il}
	\end{align}
	for $i = 1, 2, \cdots, N$, $k = 1, 2, \cdots, M_i$ and $l = 1, \cdots, L_i$.

	 \item {\color{black} In this step, we update the current estimates of $\mu_{1ik}$ and ${\mu}_{2il}$ with the estimates obtained from step 5 and we recompute the variables in steps 1 \& 2. } Set $\hat{\mu}_{1ik}^{'} = \hat{\mu}_{1ik}$ and $\hat{\mu}_{2il}^{'} = \hat{\mu}_{2il}$ for $i = 1, 2, \cdots, N$, $k = 1, 2, \cdots, M_i$ and $l = 1, 2, \cdots, L_i$. {\color{black} Recompute $D_{ij}$ and $\tilde{D}_{ij}$. Then recompute $\chi_{ijkl}^{(1)}$, $\chi_{ijkl}^{(0)}$ and $\tilde{a}_{ijkl}$ using (\ref{a1_ijkl_men1}), (\ref{a0_ijkl_men1}) and (\ref{a_noise_ijkl_men1}), respectively.}	
	 
	\item  {\color{black} In this step, we calculate the updated estimates of ${\sigma}_{1ik}^2$ and ${\sigma}_{2il}^2$, denoted by $\hat{\sigma}_{1ik}^2$ and $\hat{\sigma}_{2il}^2$, respectively, for $i = 1, 2, \cdots, N$, $k = 1, 2, \cdots, M_i$ and $l = 1, 2, \cdots, L_i$.} Define  $D_{\mu_1, ik} = \sum_{j=1}^P \sum_{l=1}^{L_i} \left( \chi_{ijkl}^{(1)} + \chi_{ijkl}^{(0)}\right) + \sum_{j=1}^{P_t} \sum_{l=1}^{L_i} \tilde{a}_{ijkl}$ and $D_{\mu_2, il} = \sum_{j=1}^P \sum_{k=1}^{M_i} \left( \chi_{ijkl}^{(1)} + \chi_{ijkl}^{(0)}\right) + \sum_{j=1}^{P_t} \sum_{k=1}^{M_i} \tilde{a}_{ijkl}$ for $i = 1, 2, \cdots, N$, $k = 1, 2, \cdots, M_i$  and $l = 1, \cdots, L_i$. Then compute
	
	{	\begin{align}
\hat{\sigma}_{1ik}^2 &= D_{\mu_1, ik}^{-1} \left[ \sum_{j=1}^{P_t}\sum_{l=1}^{L_i}  \tilde{a}_{ijkl}(\tilde{w}_{1ij} - \hat{\mu}_{1ik}^{'} )^2 \right.  + \sum_{j=1}^P \sum_{l=1}^{L_i}  \chi_{ijkl}^{(0)}  \left( \frac{t_{2ij} - \hat{\delta}^{'}}{\hat{\phi}^{'}} - \hat{d}_i^{'} - t_{1ij} - \hat{\mu}_{1ik}^{'}\right)^2 \nonumber \\
& \left. + \chi_{ijkl}^{(1)}  \left( \frac{t_{2ij} - \hat{\delta}^{'}}{\hat{\phi}^{'}} - \hat{d}_i^{'} - \hat{\tau}_i^{'} - t_{1ij} - \hat{\mu}_{1ik}^{'}\right)^2   \right], \label{sigma_1i}
	\end{align}	}
	and
	{
		\begin{align}
	\hat{\sigma}_{2il}^2 &=  D_{\mu_2, il}^{-1} \left[ \sum_{j=1}^{P_t}\sum_{k=1}^{M_i}  \tilde{a}_{ijkl}(\tilde{w}_{2ij} - \hat{\mu}_{2il}^{'} )^2  \right. + \left.\sum_{j=1}^P \sum_{k=1}^{M_i}  \left(\chi_{ijkl}^{(1)} + \chi_{ijkl}^{(0)}\right) \left( t_{4ij} - \hat{d}_i^{'} + \frac{\hat{\delta}^{'} - t_{3ij}}{\hat{\phi}^{'}} - \hat{\mu}_{2il}^{'}\right)^2\right] \label{sigma_2i}
	\end{align}}	
	for $i = 1, 2, \cdots, N$, $k = 1, 2, \cdots, M_i$  and $l = 1, \cdots, L_i$.
	
	\item {\color{black} In this step, we update the current estimates of ${\sigma}_{1ik}^{2}$ and ${\sigma}_{2il}^{2}$ with the estimates obtained from step 7 and we recompute the variables in steps 1 \& 2. } Set $\hat{\sigma}_{1ik}^{'2} = \hat{\sigma}_{1ik}^2$ and $\hat{\sigma}_{2il}^{'2} = \hat{\sigma}_{2il}^2$ for $i = 1, 2, \cdots, N$, $k = 1, 2, \cdots, M_i$ and $l = 1, 2, \cdots, L_i$. {\color{black} Recompute $D_{ij}$ and $\tilde{D}_{ij}$. Then recompute $\chi_{ijkl}^{(1)}$, $\chi_{ijkl}^{(0)}$ and $\tilde{a}_{ijkl}$ using (\ref{a1_ijkl_men1}), (\ref{a0_ijkl_men1}) and (\ref{a_noise_ijkl_men1}), respectively.}

	 \item {\color{black} In this step, we calculate the updated estimates of $d_i$, denoted by $\hat{d}_i$, for the various master-slave communication paths.}  Define $D_{d, i} = \sum_{j=1}^P \sum_{k=1}^{M_i} \sum_{l=1}^{L_i} \left( \chi_{ijkl}^{(1)} + \chi_{ijkl}^{(0)}\right) \left( \frac{1}{\sigma_{1ik}^{'2}} + \frac{1}{\sigma_{2il}^{'2}}\right)$ for $i = 1, 2, \cdots, N$. Then compute
	 {\small
	\begin{align}
\hat{d}_{i} &=  D_{d, i}^{-1} \left[ \sum_{j=1}^P \sum_{k=1}^{M_i} \sum_{l=1}^{L_i} \left(  \chi_{ijkl}^{(1)}\left[ \frac{\left( \frac{\hat{\delta}^{'} - t_{3ij}}{\hat{\phi}^{'}} + t_{4ij} - \hat{\mu}_{2il}^{'}\right)}{\sigma_{2il}^{'2}} \right. \right. \right.   \left. + \frac{\left( \frac{t_{2ij} - \hat{\delta}^{'}}{\hat{\phi}^{'}} - \hat{\tau}_i^{'} - t_{1ij} - \hat{\mu}_{1ik}^{'}\right)}{\sigma_{1ik}^{'2}} \right]      \nonumber \\
	& + \chi_{ijkl}^{(0)} \left[ \frac{\left( \frac{t_{2ij} - \hat{\delta}^{'}}{\hat{\phi}^{'}} - t_{1ij} - \hat{\mu}_{1ik}^{'}\right)}{\sigma_{1ik}^{'2}} \right. \left. \left. \left.+ \frac{\left( \frac{\hat{\delta}^{'} - t_{3ij}}{\hat{\phi}^{'}} + t_{4ij} - \hat{\mu}_{2il}^{'}\right)}{\sigma_{2il}^{'2}}\right] \right)\right] \label{d_i}
	\end{align}	}
	for $i = 1, 2, \cdots, N$.
	
	 \item {\color{black} In this step, we update the current estimates of $d_i$ using the estimates obtained from step 9 and we recompute the variables in steps 1 \& 2. }Set $\hat{d}_{i}^{'} = \hat{d}_{i}$ for $i = 1, 2, \cdots, N$. {\color{black} Recompute $D_{ij}$ and $\tilde{D}_{ij}$. Then recompute $\chi_{ijkl}^{(1)}$, $\chi_{ijkl}^{(0)}$ and $\tilde{a}_{ijkl}$ using (\ref{a1_ijkl_men1}), (\ref{a0_ijkl_men1}) and (\ref{a_noise_ijkl_men1}), respectively.}	
	 
	 \item {\color{black} In this step, we calculate the updated estimates of $\tau_i$, denoted by $\hat{\tau}_i$, for the various master-slave communication paths.} Define $D_{\tau, i} = \sum_{j=1}^P \sum_{k=1}^{M_i} \sum_{l=1}^{L_i} \frac{\chi_{ijkl}^{(1)}}{\sigma_{1ik}^{'2}}$ for $i = 1, 2, \cdots, N$. Then compute
	\begin{align}
	\hat{\tau}_{i} & =  D_{\tau, i}^{-1} \left[ \sum_{j=1}^P \sum_{k=1}^{M_i} \sum_{l=1}^{L_i}   \chi_{ijkl}^{(1)} \right.  \left. \left( \frac{\left( \frac{t_{2ij} - \hat{\delta}^{'}}{\hat{\phi}^{'}} - \hat{d}^{'}_i - t_{1ij} - \hat{\mu}_{1ik}^{'}\right)}{\sigma_{1ik}^{'2}} \right)\right] \label{tau_i}
	\end{align}	
    for $i = 1, 2, \cdots, N$.

	\item {\color{black} In this step, we update the current estimates of $\tau_i$ using the estimates obtained from step 11 and we recompute the variables in steps 1 \& 2. }Set $\hat{\tau}_{i}^{'} = \hat{\tau}_{i}$ for $i = 1, 2, \cdots, N$. {\color{black} Recompute $D_{ij}$ and $\tilde{D}_{ij}$. Then recompute $\chi_{ijkl}^{(1)}$, $\chi_{ijkl}^{(0)}$ and $\tilde{a}_{ijkl}$ using (\ref{a1_ijkl_men1}), (\ref{a0_ijkl_men1}) and (\ref{a_noise_ijkl_men1}), respectively.}

	\item {\color{black} In this step, we calculate the updated estimate of the clock offset $\delta$, denoted by $\hat{\delta}$.} Define $D_{\delta} = \sum_{i=1}^N \sum_{j=1}^P \sum_{k=1}^{M_i} \sum_{l=1}^{L_i} \left( \chi_{ijkl}^{(1)} + \chi_{ijkl}^{(0)}\right)$ $\left( \frac{1}{\sigma_{1ik}^{'2}} + \frac{1}{\sigma_{2il}^{'2}}\right)$ and compute
	\begin{align}
	\hat{\delta}&  =  \hat{\phi}^{'} D_{\delta}^{-1}  \left[ \sum_{i=1}^N \sum_{j=1}^P \sum_{k=1}^{M_i} \sum_{l=1}^{L_i} \left(  \chi_{ijkl}^{(1)} \right. \right. \left[ \frac{\left( \frac{t_{2ij}}{\hat{\phi}^{'}} - \hat{d}_i^{'} - \hat{\tau}_i^{'} - t_{1ij} - \hat{\mu}_{1ik}^{'}\right)}{\sigma_{1ik}^{'2}} \right. \nonumber \\
	& \left. - \frac{\left( t_{4ij} - \frac{t_{3ij}}{\hat{\phi}^{'}} - \hat{d}_i^{'} -  \hat{\mu}_{2il}^{'}\right)}{\sigma_{2il}^{'2}}\right]      + \chi_{ijkl}^{(0)} \left[ \frac{\left( \frac{t_{2ij}}{\hat{\phi}^{'}} - \hat{d}_i^{'} - t_{1ij} - \hat{\mu}_{1ik}^{'}\right)}{\sigma_{1ik}^{'2}} \right.  \left.\left.\left. - \frac{\left( t_{4ij} - \frac{ t_{3ij}}{\hat{\phi}^{'}} - \hat{d}_i^{'}  - \hat{\mu}_{2il}^{'}\right)}{\sigma_{2il}^{'2}}\right] \right)\right]. \label{delta_upate}
	\end{align}

	\item {\color{black} In this step, we update the current estimates of $\delta$ using $\hat{\delta}$ obtained from step 13 and we recompute the variables in steps 1 \& 2. }Set $\hat{\delta}^{'} = \hat{\delta}$. {\color{black} Recompute $D_{ij}$ and $\tilde{D}_{ij}$. Then recompute $\chi_{ijkl}^{(1)}$, $\chi_{ijkl}^{(0)}$ and $\tilde{a}_{ijkl}$ using (\ref{a1_ijkl_men1}), (\ref{a0_ijkl_men1}) and (\ref{a_noise_ijkl_men1}), respectively.}	
	
	\item {\color{black} In this step, we calculate the updated estimate of the clock skew $\phi$, denoted by $\hat{\phi}$.} Define $c_{\phi} = \sum_{i=1}^N \sum_{j=1}^P \sum_{k=1}^{M_i} \sum_{l=1}^{L_i} (\chi_{ijkl}^{(1)} + \chi_{ijkl}^{(0)})  \left( \frac{(t_{2ij} - \hat{\delta}^{'})^2 }{\sigma^{'2}_{1ik}} + \frac{(\hat{\delta}^{'} - t_{3ij})^2}{\sigma^{'2}_{2il}} \right)$, $a_{\phi} = 2NP$ and $b_{\phi}$ as
	\begin{align}
	b_{\phi} & = \sum_{i=1}^N \sum_{j=1}^P \sum_{k=1}^{M_i} \sum_{l=1}^{L_i} \chi_{ijkl}^{(1)} \left[ \frac{(\hat{d}_i^{'} + \hat{\tau}_i^{'} + t_{1ij} + \hat{\mu}_{1ik}^{'})(t_{2ij} - \hat{\delta}^{'})}{\sigma^{'2}_{1ik}} -  \frac{(t_{4ij} - \hat{d}_i^{'} - \hat{\mu}^{'}_{2il})(\hat{\delta}^{'} - t_{3ij} )}{\sigma^{'2}_{2il}} \right] \nonumber \\
	&+ \chi_{ijkl}^{(0)} \left[ \frac{(\hat{d}_i^{'} + t_{1ij} + \hat{\mu}^{'}_{1ik})(t_{2ij} - \hat{\delta}^{'})}{\sigma^{'2}_{1ik}} -  \frac{(t_{4ij} - \hat{d}_i^{'} - \hat{\mu}^{'}_{2il})(\hat{\delta}^{'} - t_{3ij} )}{\sigma^{'2}_{2il}} \right]. \label{b_phi}
	\end{align}
	 Then compute $\hat{\phi} =  \frac{\sqrt{b_{\phi}^2 - 4a_{\phi}c_{\phi}} - b_{\phi}}{2a_{\phi}}$.
	
	\item {\color{black} In this step, we update the current estimates of $\phi$ using $\hat{\phi}$ obtained from step 15 and we recompute the variables in steps 1 \& 2. }Set $\hat{\phi}^{'} = \hat{\phi}$, and repeat steps $1)-16)$.

\end{enumerate}

Since the update equations in steps $1)-16)$ {\color{black} employ} the SAGE algorithm, they inherit the desirable property that the likelihood is non-decreasing {\color{black} at each iteration} \cite{Fessler}. When the algorithm converges, we  obtain the estimate of the clock skew and offset from $\bm{\Omega}_{\pi}^{'}$. Initial values for the parameters are required to begin the SAGE algorithm. A simple ad-hoc scheme to obtain the initial values of the various parameters in $\bm{\Omega}_{\pi}$ is presented in Appendix \ref{App_sec3}.  We observe from numerical results that the proposed ad-hoc initialization scheme seems to avoid {\color{black} convergence to} local minimums in the cases studied.

{\color{black}
\subsection{Computational complexity}
Let $M_{max}$ and $L_{max}$ denote the largest element of the sets $\{M_1, M_2, \cdots, M_N \}$ and $\{L_1, L_2, \cdots, L_N \}$, respectively. At every iteration of the proposed algorithm, we would require $\mathcal{O}(N(M + P_t)M_{max}L_{max})$ additions, $\mathcal{O}(N(M + P_t)M_{max}L_{max})$ multiplications and $\mathcal{O}(NMM_{max}L_{max})$ divisions, where $\mathcal{O}(.)$ represents the big-O notation, $N$ is the number of master-slave communication paths, $M$ is the total number of two-way message exchanges and $P_t$ is the length of the vector $\tilde{\bm{w}}_{ki}$. Hence, the total computational complexity of the proposed algorithm is given by $\mathcal{O}(N_{iter}N(M + P_t)M_{max}L_{max})$ additions, $\mathcal{O}(N_{iter}N(M + P_t)M_{max}L_{max})$ multiplications and $\mathcal{O}(N_{iter}NM M_{max}L_{max})$ divisions, where $N_{iter}$ is the total number of iterations required by the algorithm to converge\footnote{From our simulations, we observed that the algorithm typically converges in $8-10$ iterations.}.
}


\begin{figure}[t]
	\centering
	\begin{subfigure}[b]{0.45\columnwidth}
		\centering
		\includegraphics[height =  1.5 in, width = \columnwidth]{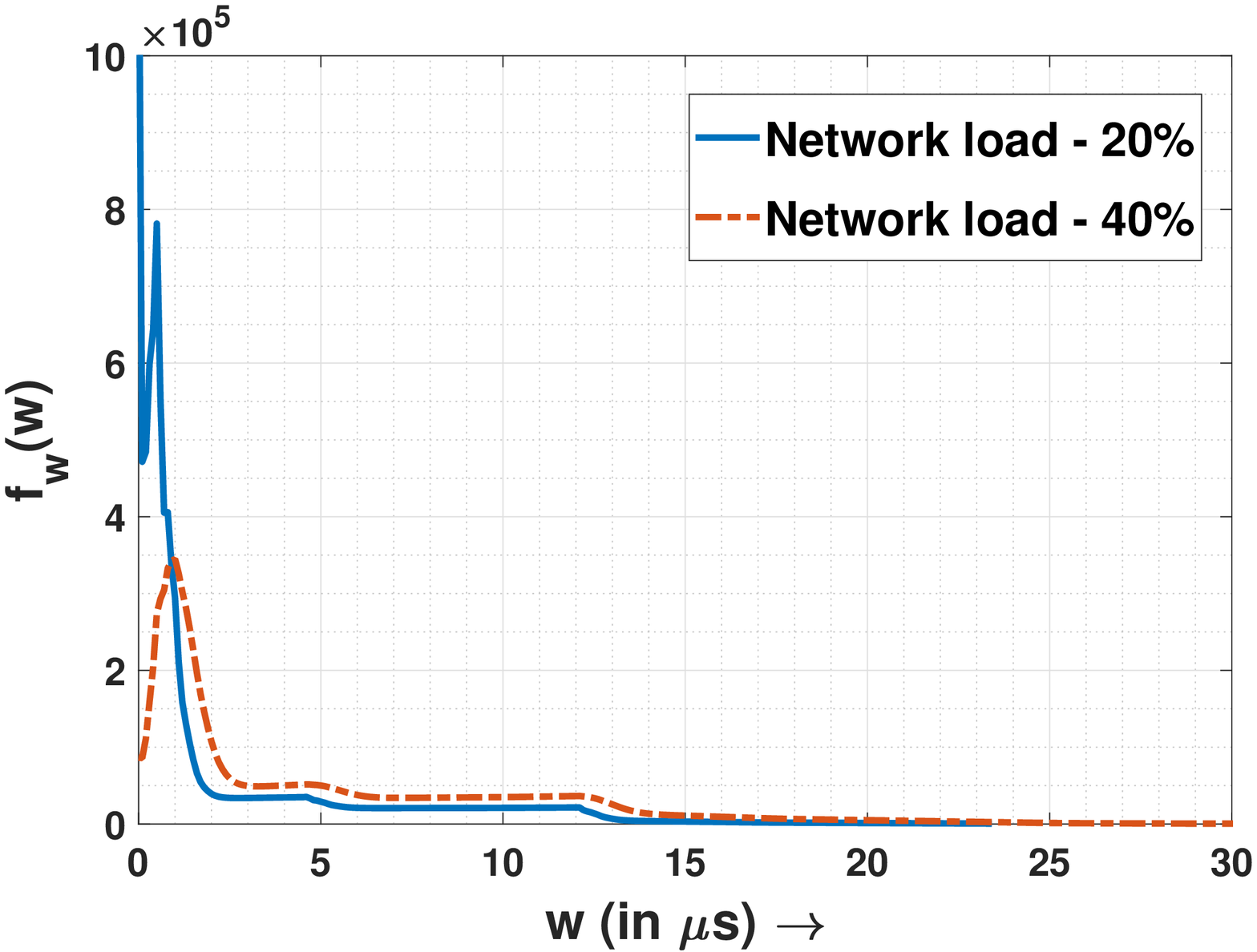}
		\caption{Empirical pdfs of random queuing delays for TM-1 under 20\% and 40\% load.}
	\end{subfigure}	
~
	\begin{subfigure}[b]{0.45\columnwidth}
		\centering
		\includegraphics[height =  1.5 in, width = \columnwidth]{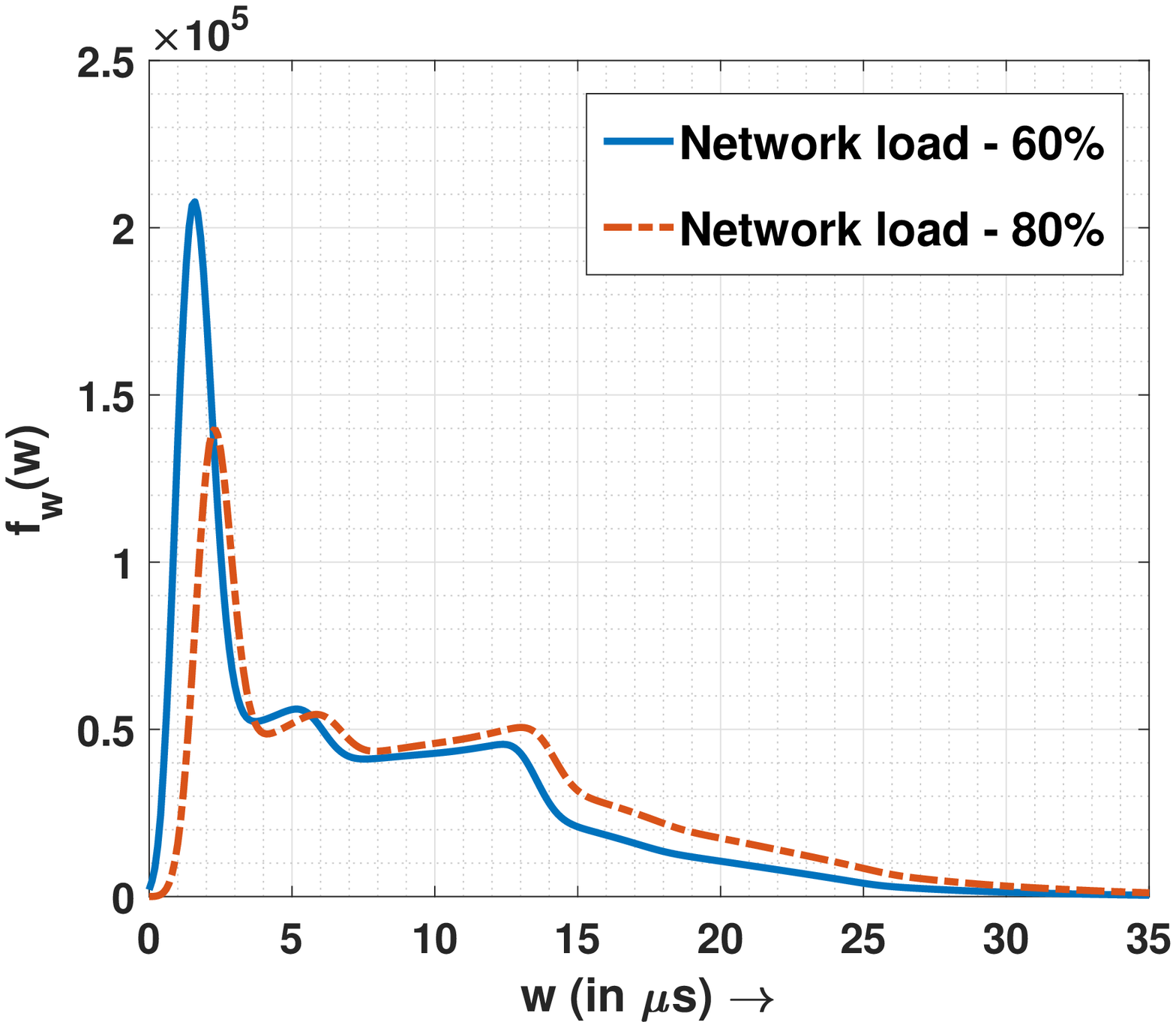}
		\caption{Empirical pdfs of random queuing delays for TM-1 under 60\% and 80\% load.}
	\end{subfigure}	
	\caption{Empirical pdfs of the random queuing delays for Traffic Model-1 under various loads.}\label{TM1_load_figures}
\end{figure}

\begin{figure}[t]
	\centering
	\begin{subfigure}[b]{0.45\columnwidth}
		\centering
		\includegraphics[height =  1.5 in, width = \columnwidth]{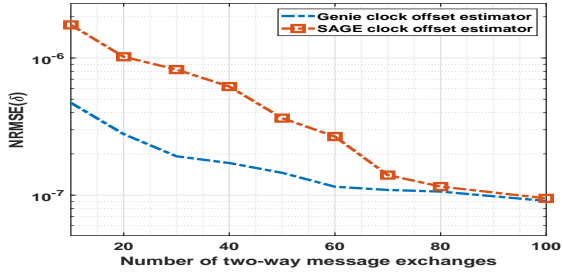}
		\caption{NRMSE of clock offset for TM-1 under 60\% load.}
	\end{subfigure}	
~
	\begin{subfigure}[b]{0.45\columnwidth}
		\centering
		\includegraphics[height =  1.5 in, width = \columnwidth]{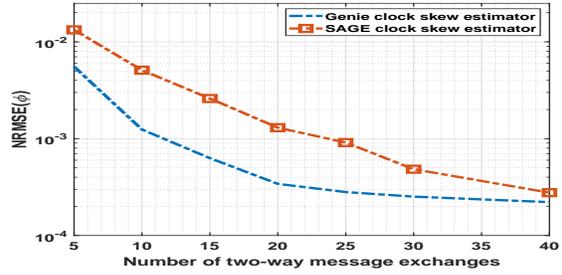}
		\caption{NRMSE of clock skew for TM-1 under 60\% load.}
	\end{subfigure}	
	\caption{NRMSE of clock skew and offset for the considered CSOE schemes under Traffic Model-1.}\label{nrmse_TM1_offset_skew_results}
\end{figure}


\begin{figure}[t]
	\centering
	\begin{subfigure}[b]{0.45\columnwidth}
		\centering
		\includegraphics[height =  1.5 in, width = \columnwidth]{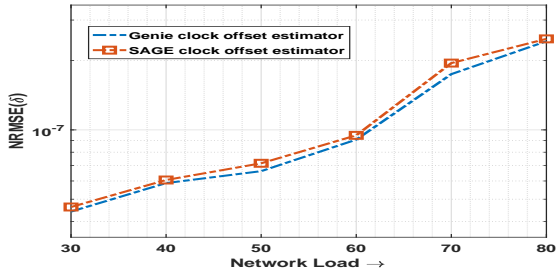}
		\caption{NRMSE of clock offset for $P = 100$. }
	\end{subfigure}	
	~
	\begin{subfigure}[b]{0.45\columnwidth}
		\centering
		\includegraphics[height =  1.5 in, width = \columnwidth]{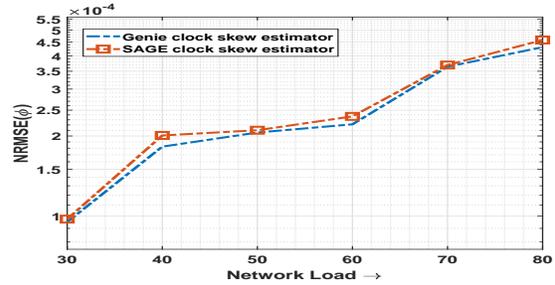}
		\caption{NRMSE of clock skew for $P = 40$. }
	\end{subfigure}	
	\caption{NRMSE of clock offset and skew for the considered CSOE schemes under Traffic Model-1 under different loads.}\label{nrmse_TM1_diffloads}
\end{figure}


\begin{figure}[t]
	\centering
	\begin{subfigure}[b]{0.45\columnwidth}
		\centering
		\includegraphics[height =  1.5 in, width = \columnwidth]{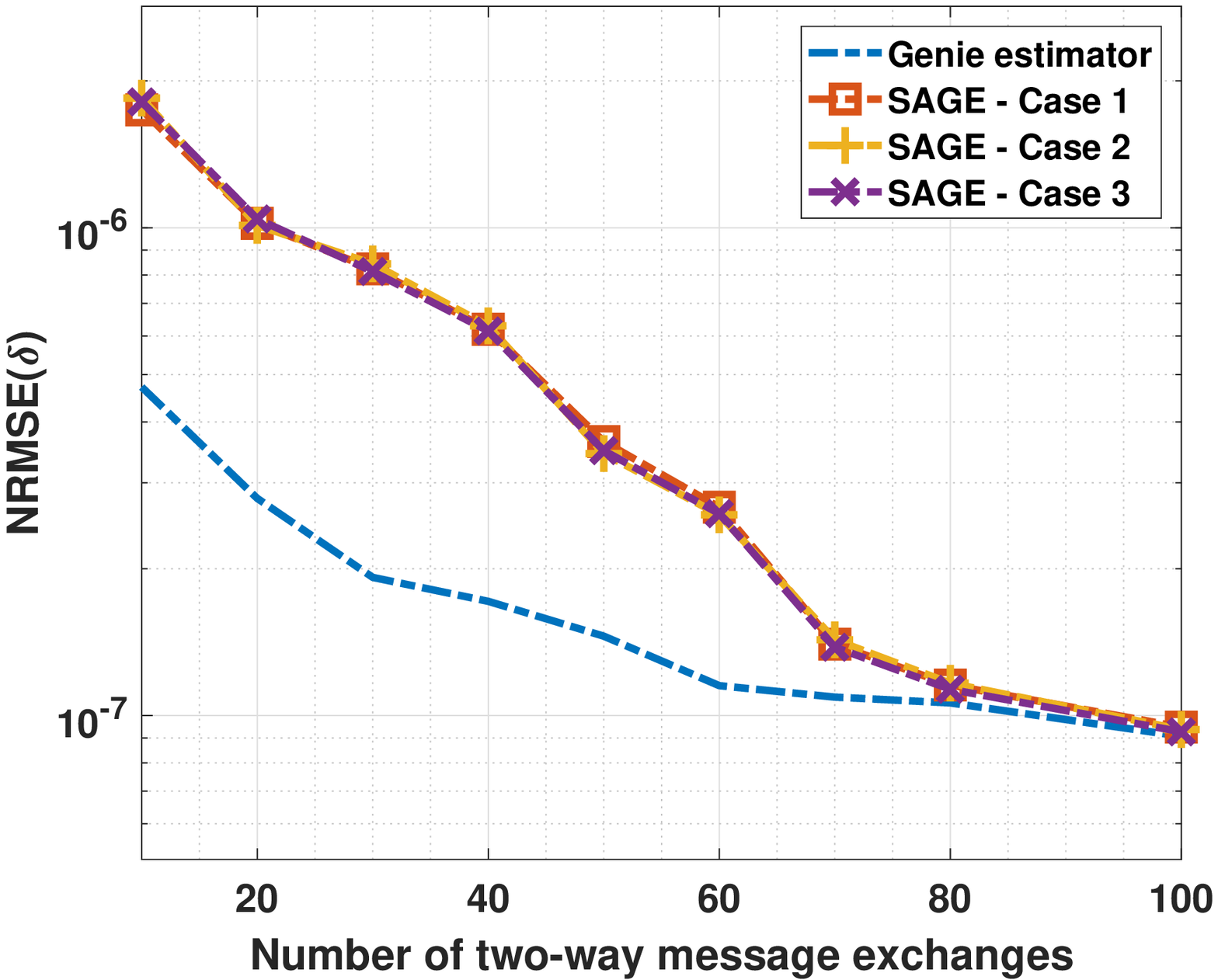}
		\caption{NRMSE of clock offset for TM-1 under 60\% load }
	\end{subfigure}	
	~
	\begin{subfigure}[b]{0.45\columnwidth}
		\centering
		\includegraphics[height =  1.5 in, width =\columnwidth]{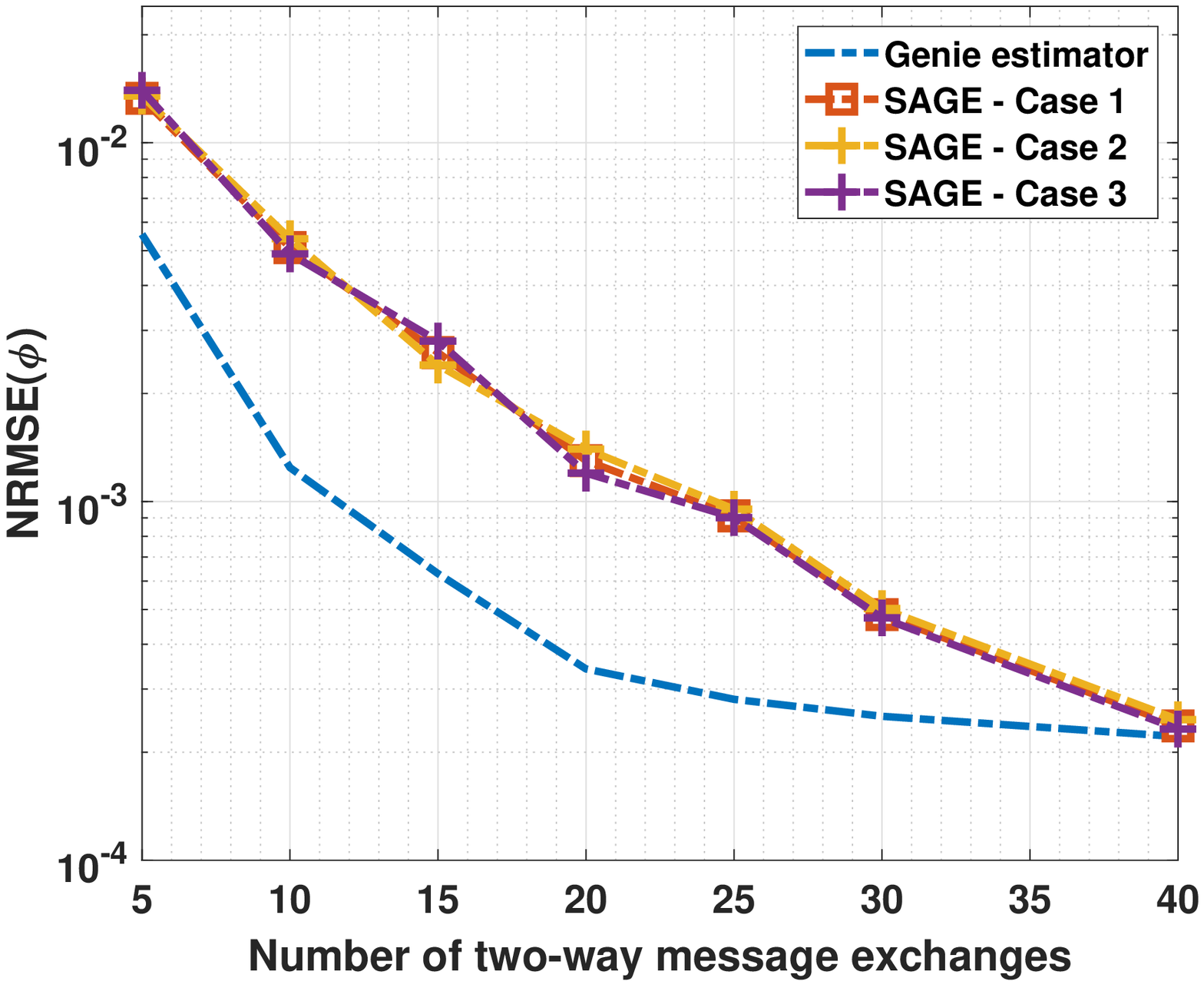}
		\caption{NRMSE of clock skew for TM-1 under 60\% load }
	\end{subfigure}	
	\caption{NRMSE of clock offset and skew for different values of $\{\phi, \delta \}$ under Traffic Model-1. We have for case 1, $\{\phi, \delta \} = \{1.01, 1~\mu s\}$, for case 2, $\{\phi, \delta \} = \{1.01, 0~\mu s\}$ and for case 3, $\{\phi, \delta \} = \{1 , 0~\mu s\}$.}\label{nrmse_TM1_diff_offset_skew_results}
	
\end{figure}

\begin{figure}[t]
	\centering
	\begin{subfigure}[b]{0.45\columnwidth}
		\centering
		\includegraphics[height =  1.5 in, width = \columnwidth]{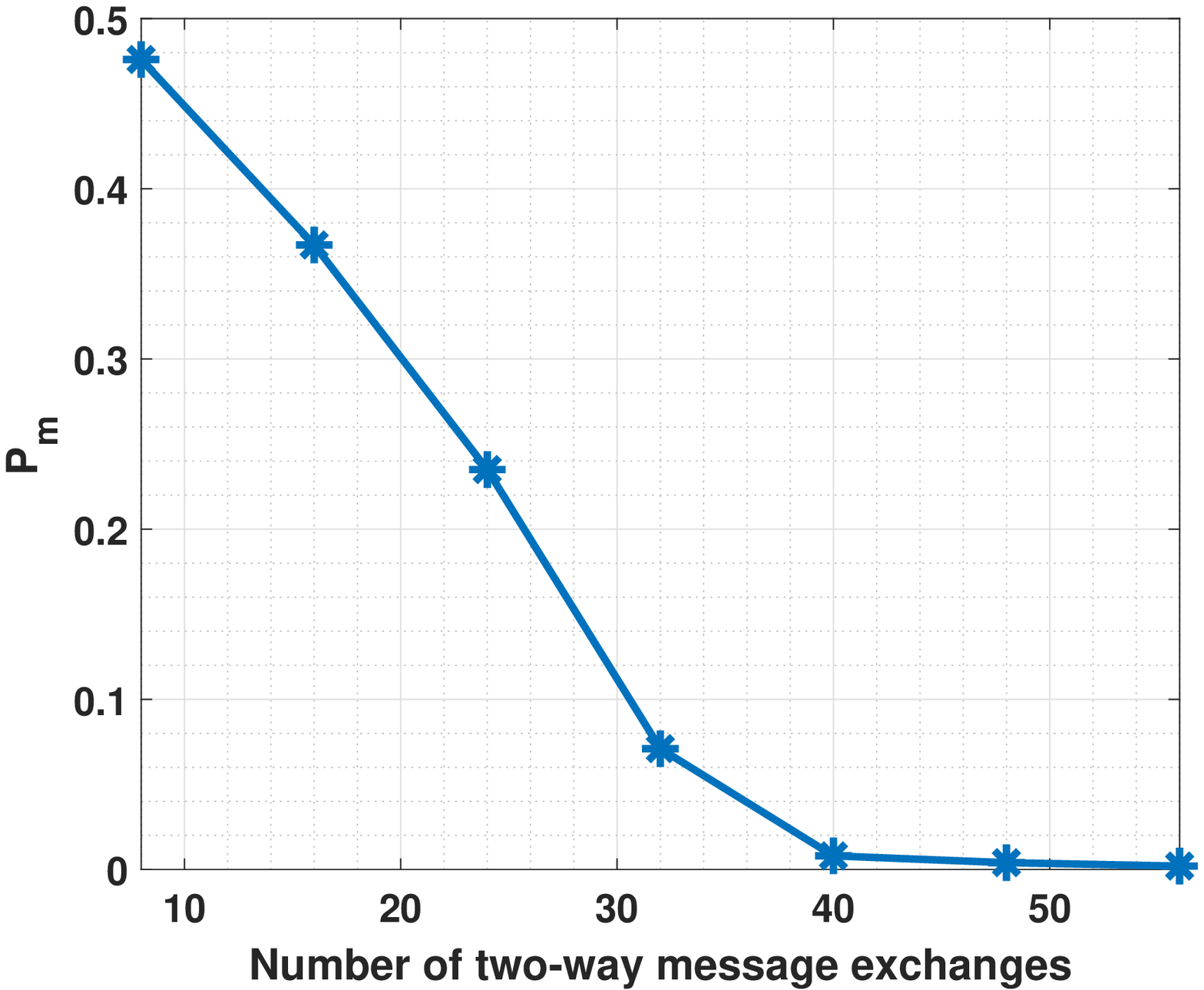}
		\caption{Probability of miss detection for TM-1 under 60\% load.  }
	\end{subfigure}	
	~
	\begin{subfigure}[b]{0.45\columnwidth}
		\centering
		\includegraphics[height =  1.5 in, width = \columnwidth]{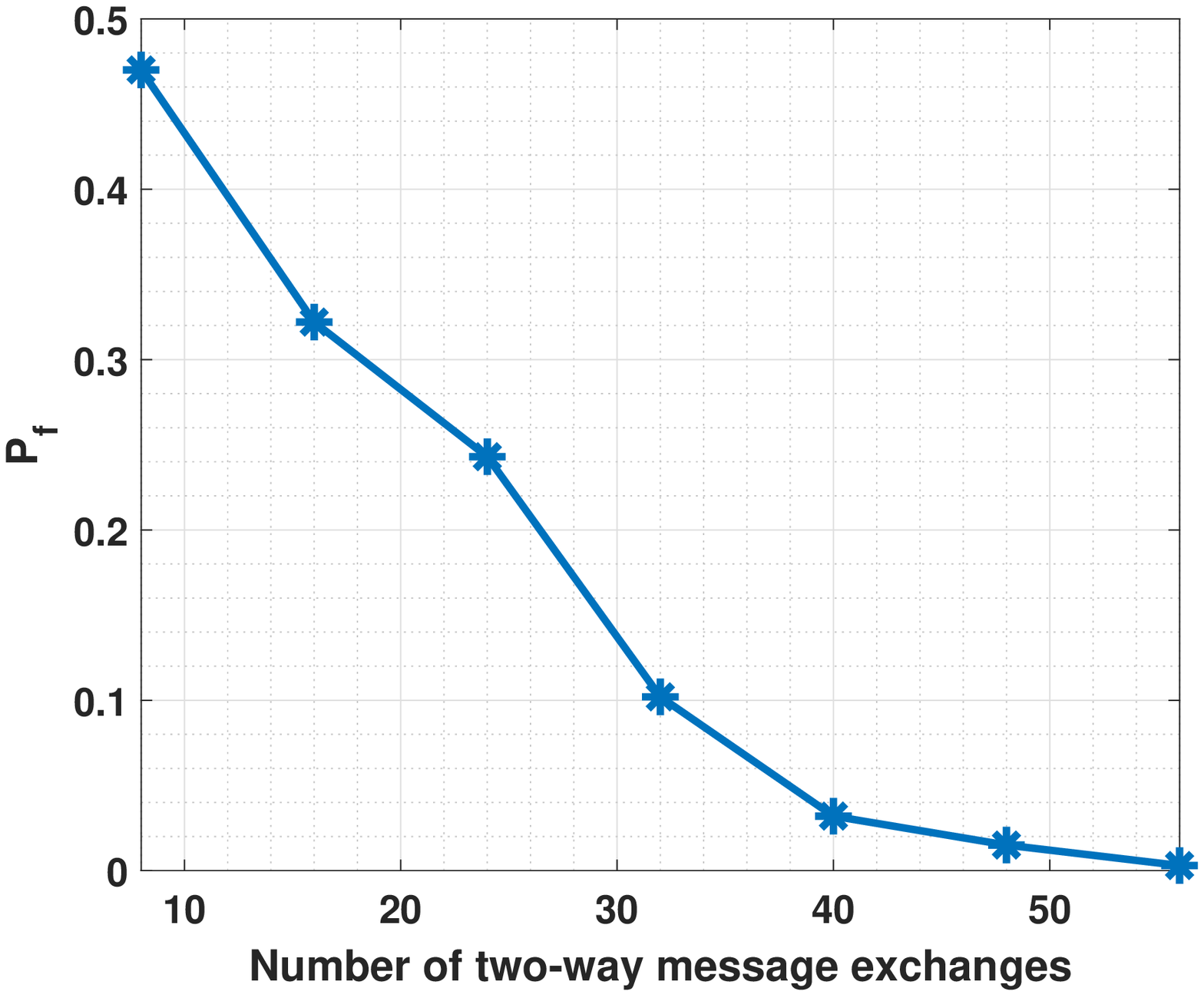}
		\caption{Probability of false alarm detection for TM-1 under 60\% load.  }
	\end{subfigure}	
	\caption{Probability of miss detection and probability of false alarm for TM-1 under 60\% load.}\label{Pm_detection}
\end{figure}

\begin{figure}[t]
	\centering
	\begin{subfigure}[b]{0.45\columnwidth}
		\centering
		\includegraphics[height =  1.5 in, width = \columnwidth]{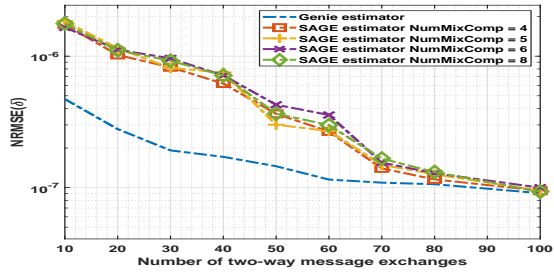}
		\caption{NRMSE of clock offset for TM-1 under 60\% load }
	\end{subfigure}	
	~
	\begin{subfigure}[b]{0.45\columnwidth}
		\centering
		\includegraphics[height =  1.5 in, width = \columnwidth]{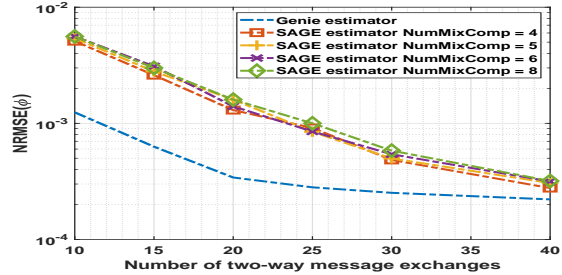}
		\caption{NRMSE of clock skew for TM-1 under 60\% load }
	\end{subfigure}	
	\caption{NRMSE of clock offset and skew for different values of number of mixing components for the Gaussian mixture model.}\label{nrmse_diffnum_mixingComp}
\end{figure}

\begin{figure}[t]
	\centering
	\begin{subfigure}[b]{0.45\columnwidth}
		\centering
		\includegraphics[height =  1.75 in, width = \columnwidth]{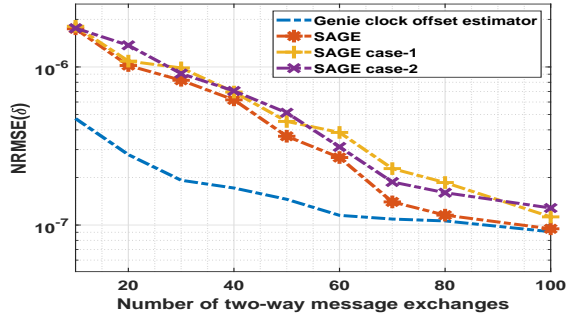}
		\caption{NRMSE of clock offset for TM-1 under 60\% load. }
	\end{subfigure}	
	~
	\begin{subfigure}[b]{0.45\columnwidth}
		\centering
		\includegraphics[height =  1.75 in, width =\columnwidth]{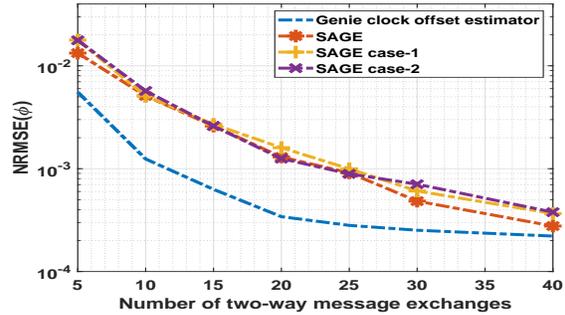}
		\caption{NRMSE of clock skew for TM-1 under 60\% load. }
	\end{subfigure}	
	\caption{  NRMSE of clock offset and skew for different cases. The samples $\bm{w}$ are assumed to be generated from a pdf corresponding to TM-1 under 60\% load. In SAGE, $\tilde{\bm{w}}_k$ is has a pdf corresponding to TM-1 under 60\% load (ideal scenario). In SAGE case-1, $\tilde{\bm{w}}_k$ is has a pdf corresponding to TM-1 under 50\% load. In SAGE case-2, $\tilde{\bm{w}}_k$ is has a pdf corresponding to TM-1 under 70\% load.}\label{nrmse_TM1_diffloads_tilde_w_offset_skew_results}
\end{figure}

\begin{figure}[t]
	\centering
	\begin{subfigure}[b]{0.45\columnwidth}
		\centering
		\includegraphics[height =  1.75 in, width = \columnwidth]{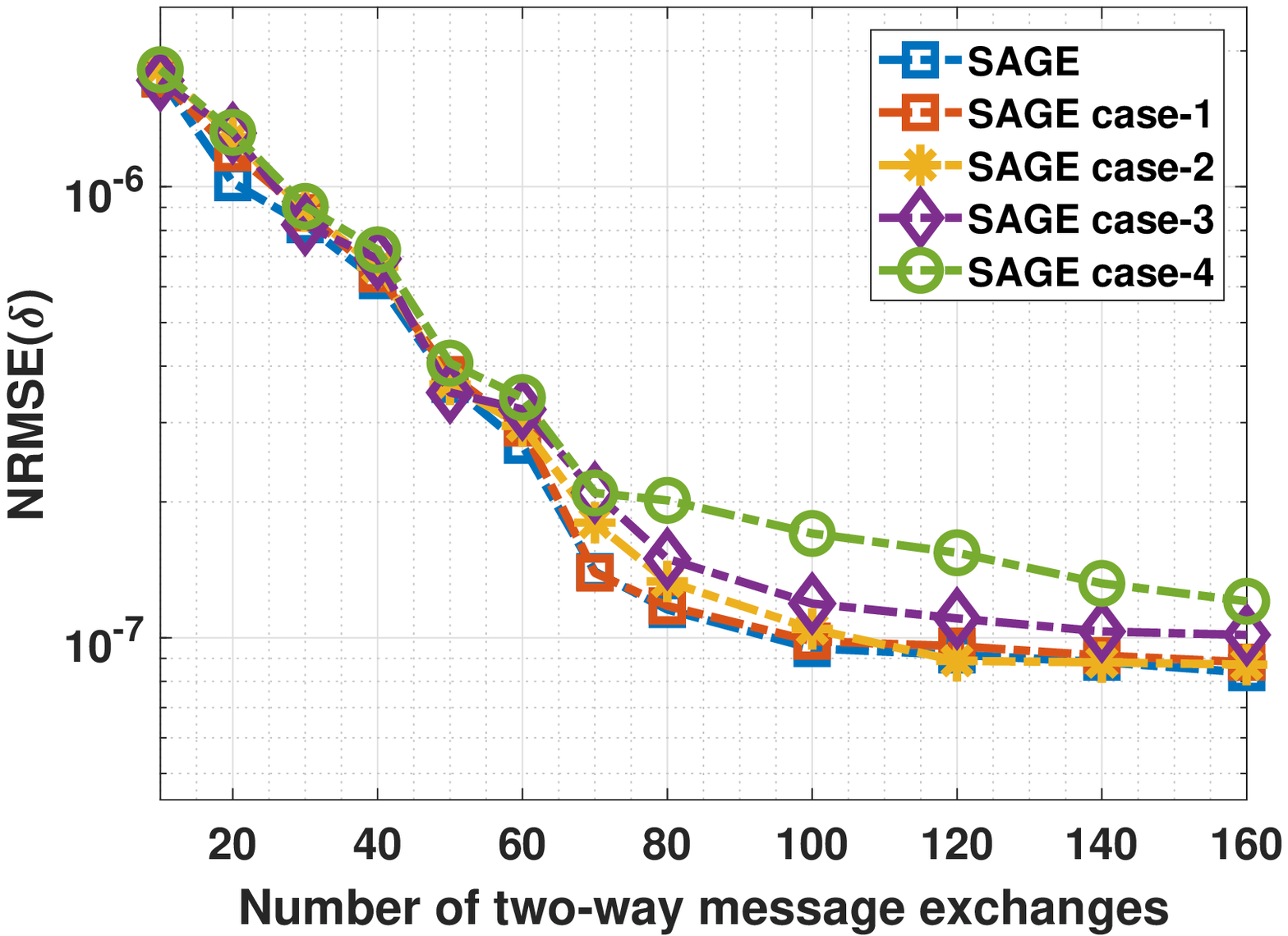}
		\caption{NRMSE of clock offset for TM-1 under 60\% load. }
	\end{subfigure}	
	~
	\begin{subfigure}[b]{0.45\columnwidth}
		\centering
		\includegraphics[height =  1.75 in, width =\columnwidth]{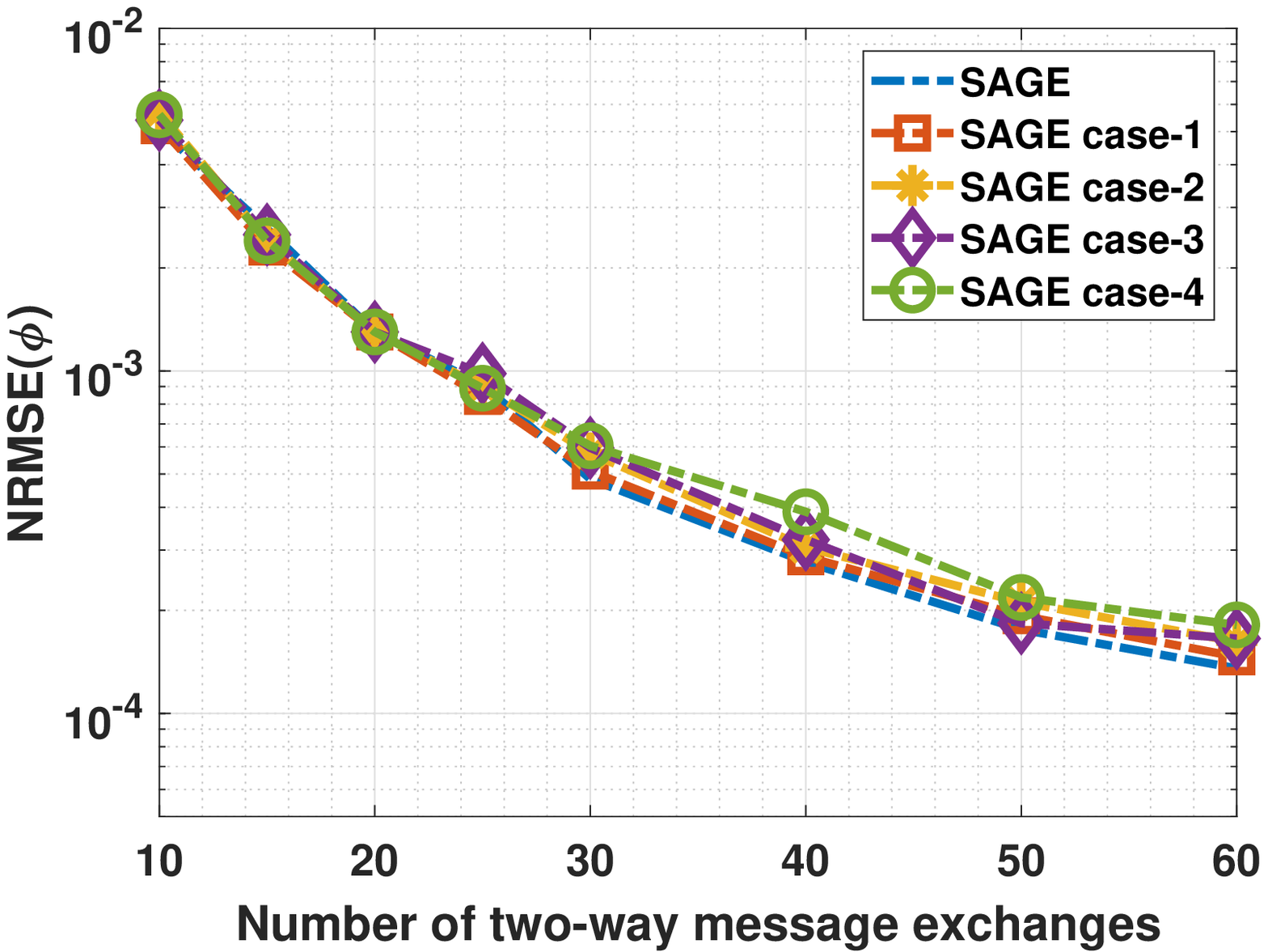}
		\caption{NRMSE of clock skew for TM-1 under 60\% load. }
	\end{subfigure}	
	\caption{  NRMSE of clock offset and skew for different cases. For SAGE, we assumed the parameters from the previous synchronization interval are estimated perfectly. In SAGE case-1, the {\color{black} common} standard deviation of the estimates of $\delta, d_i^{ms}, d_i^{sm}$ {\color{black} is} fixed { at} $1.01\times 10^{-7}$, while the standard deviation of the estimates of $\phi$ is set to $1.01\times 10^{-4}$. In SAGE case-2, the {\color{black} common} standard deviation of the estimates of $\delta, d_i^{ms}, d_i^{sm}$ is fixed {\color{black} at} $2.02\times 10^{-7}$, while the standard deviation of the estimates of $\phi$ is set to $2.02\times 10^{-4}$. In SAGE case-3, the {\color{black} common} standard deviation of the estimates of $\delta, d_i^{ms}, d_i^{sm}$ {\color{black} is} fixed {\color{black} at} $4.04\times 10^{-7}$, while the standard deviation of the estimates of $\phi$ is set to $4.04\times 10^{-4}$. In SAGE case-4, the {\color{black} common} standard deviation of the estimates of $\delta, d_i^{ms}, d_i^{sm}$ {\color{black} is} fixed at $8.08\times 10^{-7}$, while the standard deviation of the estimates of $\phi$ is set to $8.08\times 10^{-4}$.}\label{nrmse_TM1_ErrorParam_offset_skew_results}
\end{figure}

\section{Simulation Results}\label{Sec6}
In this section, we compare the performance of the proposed robust clock skew and offset estimator to the performance lower bounds via numerical simulations. We consider the LTE backhaul network scenario described in Section \ref{Sec1} for packet-swtiched networks without synchronous ethernet\footnote{In this scenario, PTP is sometimes used in conjunction with Synchronous Ethernet (SyncE) for cellular base station synchronization. Although the SyncE standards are now mature, much of the deployed base of Ethernet equipment does not support it \cite{IEEE1588v2_whitepaper}.}. PTP is the primary {\color{black} synchronization} option for operators with packet{\color{black}-switched} backhaul networks that do not support SyncE \cite{IEEE1588v2_whitepaper, IEEE1588v2_Huawei}. For simplicity, we assume $f_{{1i}}(.) = f_{{2i}}(.) = f_{{w}}(.)$ for $i = 1, 2, \cdots, N$. {\color{black} However, the proposed algorithm does not assume that all the pdfs are the same. }Further, we assume the deterministic path delays are identical across all the master-slave communication paths, i.e., $d_1 = d_2 = \cdots = d_N = d$, where $d$ denotes an unknown deterministic path delay parameter. We first briefly describe the approach used to generate the random queuing delays in our simulations.

\subsection{Generation of the random queuing delays}
We follow the approach given in \cite{Anand_2015} for generating the random queuing delays in LTE backhaul networks.  We consider a Gigabit Ethernet network consisting of a cascade of $10$ switches between the master and slave nodes. A two-class non-preemptive priority queue is used to model the traffic at each switch. The network traffic at the switch {\color{black} is comprised} of the lower priority background traffic and the higher priority synchronization messages. We assume cross-traffic flows, where new background traffic is injected at each switch and this traffic exits at the subsequent switch. The arrival times and size of background traffic packets injected at each switch are assumed to be statistically independent. We use Traffic Model 1 (TM-1) from the ITU-T specification G.8261 \cite{ITU} for generating the background traffic at each switch. The interarrival times between packets in background traffic are assumed to follow an exponential distribution, and we set the rate parameter of each exponential distribution accordingly to obtain the desired load factor, i.e., the percentage of the total capacity consumed by background traffic\cite{Anand_2015}. The empirical pdf of the PDV in the backhaul networks was obtained in \cite{Anand_2015} for different load factors and are shown in \ref{TM1_load_figures}. The timestamps $t_{1ij}$ and $t_{4ij}$ are set to $60j$ $\mu s$ and $60j$ $\mu s + 30 \mu s$, respectively, for $i = 1, \cdots, N$ and $j = 0, 1, \cdots, P-1$. For a given value of parameters, the timestamps $t_{2ij}$ and $t_{3ij}$ are then generated using (\ref{timestamps}).

\subsection{Numerical results}
In our results, we use the $\mbox{NRMSE}(\hat{\delta}) = \sqrt{\mbox{NMSE}(\hat{\delta})}$ and $\mbox{NRMSE}(\hat{\phi}) = \sqrt{\mbox{NMSE}(\hat{\phi})}$ metrics defined in (\ref{NRMSE_offset}) for evaluating the performance of the considered CSOE schemes. We evaluate the NRMSE performance of the clock skew and offset estimate obtained from the SAGE-CSOE scheme described in Section \ref{Sec4} and compare it against the NRMSE performance of the genie optimum estimator of $\delta$ and $\phi$ calculated using (\ref{Minimax_offset}) and (\ref{Minimax_skew}), respectively. {\color{black} In our numerical results presented in Figures 2-7, we approximate the multidimensional integrals in (\ref{Minimax_offset}) and (\ref{Minimax_skew}) with Riemann summations. We approximate the integral over $\mathbb{R}^+$ (corresponding to $\phi$) using Riemann sums by setting the width of the Riemann summation bins to $0.001$ and the limits of the integral to $[0.5, 2]$ and the integral over $\mathbb{R}$ (corresponding to $\delta$, $d_i$ and $\tau_i$) is approximated using Riemann sums by setting the width of the Riemann summation bins to $0.01$ $\mu s$ and the limits of the integral to $[-10 \mu s, 10 \mu s]$. We tried smaller bin-widths and observed from the results that the calculation were quite accurate.} We assume the availability of $N = 3$ master-slave communication paths with one path having an unknown asymmetry between the deterministic path delays. The values of $\phi$ and $\delta$ are fixed to $1.01$ and $1~\mu s$, respectively. The value of $d$ is set as $1~\mu s$, i.e., $d_1 = d_2 = d_3 = d = 1~\mu s$. For the master-slave communication path with an unknown asymmetry, we set the value of $\tau_i$ to $4~\mu s$. The user-defined parameter $d_{\tau}$ is set to $2~\mu s$. The number of mixture components used in the GMM approximation is set to $4$, {\color{black} and the value of $P_t$ is fixed as $P$, where $P$ is the number of two-way message exchanges used in the calculation of $\delta$ and $\phi$.} Some key observations from the results are listed below:

\begin{enumerate}
	\item \emph{Performance of the robust CSOE schemes:} The NRMSE performance of the proposed robust iterative CSOE scheme against the NRMSE of the optimum estimator is presented in Figures \ref{nrmse_TM1_offset_skew_results} and \ref{nrmse_TM1_diffloads}. In Figure \ref{nrmse_TM1_offset_skew_results}, we observe that the performance of the robust iterative SAGE-CSOE scheme improves with an increase in the number of two-way message exchanges, $P$, and exhibits performance close to the genie optimum estimator for a sufficiently large number of two-way message exchanges.  {\color{black} As expected,} the optimum estimators exhibits the {\color{black} smallest} NRMSE {\color{black} due to} prior information on which of the master-slave communication paths have unknown deterministic path asymmetry as well as the complete information regarding $f_w(.)$.  In Figure \ref{nrmse_TM1_diffloads}, we evaluate the performance of the robust scheme for TM-1 under different loads for a fixed value of $P$. We observe that the proposed robust clock skew and offset estimation scheme exhibits a performance close to the NRMSE of the optimum estimators for various network scenarios.
	
	
    \item \emph{Performance comparison for different values of clock skew and offset:} Figure \ref{nrmse_TM1_diff_offset_skew_results} shows us the performance of the robust iterative SAGE-CSOE scheme for different values of $\phi$ and $\delta$. The performance lower bounds from Proposition \ref{Minimax_optimum_estimator_phase_freq} are independent of the parameter values as is any invariant estimation scheme (see Chapter 6, \cite{Berger}). From the results, we also observe that the NRMSE performance of the SAGE-CSOE {\color{black} appears to be nearly} independent of the parameter values {\color{black} in the cases shown.}
    
    {\color{black}
    \item \emph{Probability of miss detection ($P_m$) and probability of false alarm ($P_f$):}  Let $\hat{\pi}_{i}^{(s)}$ denote the estimate of $\pi_i$ (for $i = 1, 2, \cdots, N$) obtained from the SAGE algorithm after convergence. The $i^{\mbox{th}}$ master-slave communication path is declared as asymmetric if $\hat{\pi}_{i}^{(s)} \ge 0.5$, else the $i^{\mbox{th}}$ master-slave communication path is declared as symmetric. We define the probability of miss detection (denoted by $P_m$) as the probability of identifying an asymmetric master-slave communication path as being symmetric, while the probability of false alarm (denoted by $P_f$) is defined as the probability of identifying a symmetric master-slave communication path as  asymmetric. Figure \ref{Pm_detection} shows us the $P_m$ and $P_f$ for the SAGE-CSOE scheme for different values of two-way message exchanges. As expected, $P_m$ and $P_f$ decrease to $0$ as we increase the number of two-way messages exchanges used.
    }     
    
    {\color{black} 
    \item \emph{Performance comparison for different values of number of GMM components:} Figure \ref{nrmse_diffnum_mixingComp} shows us the performance of the proposed SAGE algorithm for a different number of mixing components in the GMM\footnote{The number of mixing  components is assumed to be the same for all the master-slave communication paths in the forward and reverse paths, i.e. $M_1 = M_2 = \cdots = M_N = L_1 = L_2 = \cdots = L_N$.}. As seen from the results, there is no noticeable degradation in the performance of the SAGE CSOE scheme, possibly indicating that the performance of the algorithm is relatively robust against over-fitting.}
   
  { \color{black}
    \item \emph{Performance when the pdf of $\tilde{\bm{w}}_k$ is different than the pdf of $\bm{w}_k$: } In certain scenarios, the pdf of $\tilde{\bm{w}}_k$ for $k = 1, 2$ may be slightly different from the pdf of $\bm{w}_k$ for $k = 1, 2$. This could be a result of some  network conditions slowly changing with time across different blocks of two-way message exchanges 
    (what we called windows). Figure \ref{nrmse_TM1_diffloads_tilde_w_offset_skew_results} shows us the performance of the proposed algorithm for this scenario. We observe a slight degradation in the performance of the algorithm compared to the scenario when the pdf of $\tilde{\bm{w}}_k$ is identical to $\bm{w}_k$. However, as the number of two-way message exchanges is increased, the performance of the algorithm improves indicating that the algorithm is relatively robust against slowly changing network conditions with a sufficient number of two-way exchanges.


    \item \emph{Inaccurate previous synchronization parameters:} We now consider the scenario where $\phi$, $d_i^{ms}$, $d_i^{sm}$ and $\delta$ from the previous synchronization window are inaccurate and model them as Gaussian random variables with the mean being the true values of $1.01$, $1~\mu s$, $1~\mu s$ and $1~\mu s$, respectively. The common standard deviation of $d_i^{ms}$, $d_i^{sm}$ and $\delta$ from the last synchronization is varied from $1.01 \times 10^{-7}$ to $8.08 \times10^{-7}$, while the standard deviation of $\phi$ is varied from $1.01 \times 10^{-4}$ to $8.08\times10^{-4}$. Figure \ref{nrmse_TM1_ErrorParam_offset_skew_results} shows the performance of the proposed CSOE scheme. From the results, we observe a noticeable degradation in the performance of the robust CSOE scheme, especially for larger values of the standard deviations of $\phi$, $d_i^{ms}$, $d_i^{sm}$ and $\delta$ from the previous synchronization window. However, the performance improves with an increasing number of two-way message exchanges and is close to the case where the parameters are known perfectly for a sufficiently large number of message exchanges.}

\end{enumerate}

\section{Conclusion}
{\color{black} In this paper, assuming the availability of multiple master-slave communication paths, we have developed useful lower bounds on the skew normalized mean square estimation error for a clock skew and offset estimation  scheme in the presence of unknown path asymmetries. Also, we developed a robust iterative clock skew and offset estimation scheme that employs the SAGE algorithm for jointly estimating the clock skew and offset.  The robust iterative clock skew and offset estimation scheme has low computational complexity and does not require the complete information regarding the statistical distributions of the queuing delays. The robust scheme exhibits a skew normalized mean square estimation error close to our performance lower bounds in several network scenarios. Furthermore, a number of time synchronization protocols including NTP\cite{NTP}, TPSN \cite{TPSN},  LTS \cite{LTS}, and RBS \cite{RBS} are built on message exchanges. The proposed robust iterative scheme can be easily modified for these protocols.}

\appendices
\section{Proof of Proposition \ref{Minimax_optimum_estimator_phase_freq}}\label{App_sec1}
\noindent
\begin{proof}	
	Here we present the beautiful and complicated invariant decision theory from \cite{Berger, Lehmann} in a simple way to present our proof. In \cite{Berger, Lehmann}, it was shown that the right invariant prior, $\pi^r(.)$, on ${\mathcal{G}}_{M}$ from (\ref{M_model_group}) is obtained by finding the function that satisfies $\int_{\mathcal{A}} \pi^r(\bm{\theta}) d\bm{\theta} = \int_{\mathcal{A}_0^{(r)}} \pi^r(\bm{\theta}_{0}^{(r)}) d\bm{\theta}_{0}^{(r)}$
	for all $\mathcal{A} \subseteq \bm{\Theta}$, for all $\bar{g}_{\phi, \bm{\gamma}, {\delta}}(.) \in \bar{\mathcal{G}}_{M}$ and for all $\bm{\theta}_0 = (\phi_0, \bm{\gamma}_0, {\delta}_0) \in \bm{\Theta}$. The variables $\mathcal{A}_{0}^{(r)}$ and $\bm{\theta}_{0}^{(r)}$ are defined as follows:
	\begin{align}
	\mathcal{A}_{0}^{(r)} & =  \{ \bm{\theta}_{0}^{(r)} = \left(\phi_{0}^{(r)}, \bm{\gamma}_{0}^{(r)}, {\delta}_{0}^{(r)}\right) :   \bm{\theta}_{0}^{(r)} =  \bar{g}_{\phi, \bm{\gamma}, {\delta}}(\bm{\theta}_0), (\phi, \bm{\gamma}, {\delta}) \in \mathcal{A} \}, \\
	& =  \{ \bm{\theta}_{0}^{(r)} = (\phi\phi_0, \bm{\gamma}_0+\bm{\gamma}/\phi_0, \phi{\delta}_0 + {\delta})  : (\phi, \bm{\gamma}, {\delta}) \in \mathcal{A} \}. \label{M_model_transform}
	\end{align}
	

	 The right invariant prior for $\bar{\mathcal{G}}_{M}$ is given by $\pi^r(\bm{\theta}) =  \phi^{N+K-1}$. To see this, note that {\color{black} (from change of variables)}
	\begin{align}
	\int_{\mathcal{A}} \phi^{N+K-1} d\bm{\theta} & = \int_{\mathcal{A}_{0}^{(r)}} \left(\frac{\phi_{0}^{(r)}}{\phi_0}\right)^{N+K-1} \frac{d\bm{\theta}}{d\bm{\theta}_{0}^{(r)}} d\bm{\theta}_{0}^{(r)} = \int_{\mathcal{A}_{0}^{(r)}}  (\phi_{0}^{(r)})^{N+K-1} d\bm{\theta}_{0}^{(r)},
	\end{align}
	since the Jacobian of the transformation in (\ref{M_model_transform}) is given by
	\begin{align}
	\frac{d\bm{\theta}_{0}^{(r)}}{d\bm{\theta}} & =  \det\left(\begin{bmatrix}
	\frac{\partial \phi_{0}^{(r)}}{\partial \phi} & \frac{\partial \phi_{0}^{(r)}}{\partial \bm{\gamma}}  & \frac{\partial \phi_{0}^{(r)}}{\partial {\delta}}   \\
	\frac{\partial \bm{\gamma}_{0}^{(r)}}{\partial \phi} & \frac{\partial \bm{\gamma}_{0}^{(r)}}{\partial \bm{\gamma}}  & \frac{\partial \bm{\gamma}_{0}^{(r)}}{\partial {\delta}}  \\
	\frac{\partial {\delta}_{0}^{(r)}}{\partial \phi} & \frac{\partial {\delta}_{0}^{(r)}}{\partial \bm{\gamma}}  & \frac{\partial {\delta}_{0}^{(r)}}{\partial {\delta}}  \\
	\end{bmatrix}\right)  = \det\left(\begin{bmatrix}
	\phi_0 & \bm{0}_{N+K}^T & 0 \\
	\bm{0}_{N+K} & \frac{1}{\phi_0} \bm{I}_{N+K} & \bm{0}_{N+K} \\
	{\delta}_0 & \bm{0}_{N+K}^T & 1
	\end{bmatrix}\right)  = \frac{1}{\phi_0^{N+K-1}}. \nonumber
	\end{align}
	
	 {\color{black}A optimum invariant estimator of $\delta$} under $\mathcal{G}_{M}$ from (\ref{M_model_group}), denoted by $\hat{\delta}_{opt}$, can now be obtained by solving  \cite{Berger, Lehmann}
\begin{eqnarray}\label{Bayes_risk_Mmodel}
\hat{\delta}_{opt}(\bm{y})  =  \operatorname*{\arg \min}_{\hat{\delta}} \int_{\bm{\Theta}} \frac{(\hat{\delta}(\bm{y}) - \delta)^2}{\phi^2}  \pi^r(\bm{\theta}|\bm{y}) d\bm{\theta},
\end{eqnarray}
where $\pi^r(\bm{\theta}|\bm{y}) =  \frac{f(\bm{y}|\bm{\theta})\pi^r(\bm{\theta})}{\int_{\bm{\Theta}}f(\bm{y}|\bm{\theta})\pi^r(\bm{\theta}) d\bm{\theta}}$ and $\pi^r(\bm{\theta})$ is the right invariant prior corresponding to $\bar{\mathcal{G}}_{M}$\footnote{We should mention here that right invariant prior, $\pi^r(.)$ need not be an actual probability density function \cite{Berger}}. To find $\hat{\delta}_{opt}$, we differentiate the objective function in (\ref{Bayes_risk_Mmodel}) with respect to $\hat{\delta}(\bm{y})$, set the result equal to zero and solve for $\hat{\delta}(\bm{y})$. We have
\begin{align}
\hat{\delta}_{opt}(\bm{y}) & =  \frac{\int_{\bm{\Theta}} \frac{\delta}{\phi^2} \pi^r(\bm{\theta}|\bm{y}) d\bm{\theta} }{\int_{\bm{\Theta}} \frac{1}{\phi^2} \pi^r(\bm{\theta}|\bm{y}) d\bm{\theta}} 
& =  \frac{\int_{\bm{\Theta}} \frac{\delta}{\phi^2} f(\bm{y}|\bm{\theta})\pi^r(\bm{\theta}) d\bm{\theta} }{\int_{\bm{\Theta}} \frac{1}{\phi^2} f(\bm{y}|\bm{\theta})\pi^r(\bm{\theta}) d\bm{\theta}} = \frac{ \int_{\mathbb{R}^+}\int_{\mathbb{R}^{N+K+1}} \frac{\delta \Gamma_1(\phi, \delta, \bm{d}, \bm{\tau}, \bm{y}) \Gamma_0(\phi, \delta, \bm{d}, \bm{y})}{\phi^{2NP - N - K + 3}} d\bm{\tau} d(\bm{d}) d\delta d\phi }{ \int_{\mathbb{R}^+}\int_{\mathbb{R}^{N+K+1} } \frac{\Gamma_1(\phi, \delta, \bm{d}, \bm{\tau}, \bm{y}) \Gamma_0(\phi, \delta, \bm{d}, \bm{y})}{\phi^{2NP - N - K + 3}}   d\bm{\tau} d(\bm{d}) d\delta d\phi}, 
\end{align}
where $\Gamma_1(\phi, \delta, \bm{d}, \bm{\tau}, \bm{y})$ and $\Gamma_0(\phi, \delta, \bm{d}, \bm{y})$ are defined in Proposition \ref{Minimax_optimum_estimator_phase_freq}. 
{\color{black} Using a similar derivative-based approach, we obtain $\hat{\phi}_{opt}(\bm{y})$ defined in Proposition 1.}
\end{proof}

\section{Outline of Derivation of Update Equations}\label{App_sec2}
{\color{black}
	The first step in the EM algorithm is the specification of a set of ``complete data" $\bm{X}_c$ and ``incomplete data" $\bm{X}$ for the problem \cite{EM_mixtures, EM_main_article}. The pdf's for $\bm{X}$ and $\bm{X}_c$ are characterized by a set of common parameters $\bm{\Phi}$. The complete data is not available, but it is chosen in such a way so that if it were available, then the MLE of $\bm{\Phi}$ would be easy to find. The EM algorithm addresses this situation and provides an iterative procedure for the maximum likelihood estimation of $\bm{\Phi}$ based on the incomplete data $\bm{X}$. The SAGE algorithm \cite{Fessler} is closely related to the EM algorithm, except that the parameter set is partitioned into subsets $\bm{\Phi}_1, \bm{\Phi}_2, \cdots, \bm{\Phi}_M$ with $\bm{\Phi} = \bm{\Phi}_1 \bigcup \bm{\Phi}_2 \cdots \bigcup \bm{\Phi}_M$. Then, on each iteration, $\bm{\Phi}_1$ is updated with $\bm{\Phi}_2, \cdots, \bm{\Phi}_M $ fixed, followed by the update of $\bm{\Phi}_2$ with $\bm{\Phi}_1, \cdots, \bm{\Phi}_M $ fixed and so on. The sequence of estimates produced by the SAGE algorithm has non-decreasing likelihood for the incomplete data \cite{Fessler}. In this paper, we apply the SAGE algorithm to the considered problem. The SAGE algorithm update equations are derived as follows. The parameter set to be estimated is $\bm{\Omega}_{\pi} = [\phi, \delta, d_1, \cdots, d_N, \tau_1, \cdots, \tau_N, \bm{\pi}, \bm{\alpha}_1, \cdots, \allowbreak \bm{\alpha}_N, \bm{\beta}_1, \cdots, \bm{\beta}_N, \bm{\mu}_{11}, \allowbreak \cdots, \bm{\mu}_{1N}, \bm{\sigma}_{11}, \cdots, \bm{\sigma}_{1N}, \allowbreak \bm{\mu}_{21}, \cdots, \bm{\mu}_{2N}, \bm{\sigma}_{21}, \allowbreak \cdots, \bm{\sigma}_{2N}]$. The number of mixture components in the forward and reverse path, denoted by $M_i$ and $L_i$, respectively, are assumed to be fixed for $i = 1, 2, \cdots, N$. The incomplete data set $\bm{X}$ consists of the observed timestamps
\begin{align}
\bm{X} & = \{ t_{1ij}, t_{2ij}, t_{3ij}, t_{4ij}, \tilde{w}_{1ij_t}, \tilde{w}_{2ij_t} : i = 1, 2, \cdots, N, j = 1, 2, \cdots, P, j_t = 1, 2, \cdots, P_t\}
\end{align}
from (\ref{timestamps}). The complete data set, denoted by $\bm{X}_c$, is defined as
\begin{align}
\bm{X}_c & = \{ t_{1ij}, (t_{2ij}, z_{ij}, r_{ij}), (t_{3ij}, s_{ij}), t_{4ij}, (\tilde{w}_{1ij_t}, \tilde{r}_{ij}),  (\tilde{w}_{2ij_t}, \tilde{s}_{ij}) : i = 1, 2, \cdots, N, j = 1, 2, \cdots, P, \nonumber \\
& j_t = 1, 2, \cdots, P_t\}
\end{align}
where $z_{ij} \in \{0, 1\}$ identifies whether the $j^{th}$ two-way message exchange at the $i^{th}$ path has an unknown path asymmetry, $r_{ij} \in \{1, 2, \cdots, M_i\}$ identifies which term in the {\color{black} mixture pdf approximation of $f_{1i}(.)$ in (\ref{fwd_mixture})} produced the random queuing sample in the forward path time stamps $t_{2ij}$, and $s_{ij} \in \{1, 2, \cdots, L_i\}$ identifies which term in the mixture pdf approximation of $f_{2i}(.)$ in (\ref{fwd_mixture}) produced the random queuing sample in the reverse path time stamps $t_{3ij}$. Similarly, $\tilde{r}_{ij} \in \{1, 2, \cdots, M_i\}$ and $\tilde{s}_{ij} \in \{1, 2, \cdots, L_i\}$ identifies which term in the mixture pdf produced the random queuing samples $\tilde{w}_{1ij}$ and $\tilde{w}_{2ij}$, respectively. The definition of the complete data for mixture models is discussed in \cite{EM_mixtures}. The incomplete data log likelihood is given in (\ref{Incomplete_logLL}) and the complete data log likelihood, denoted by  $\mathcal{L}_{com}({\bm{\Omega}}_{\pi}|\bm{X}_c)$, is defined in (\ref{Complete_logLL}) as
\begin{align}
 & \sum_{i=1}^N \sum_{j=1}^P z_{ij}\ln \left[\pi_i\alpha_{r_{ij}} \mathcal{P}_{\mu_{1r_{ij}}, \sigma_{1r_{ij}}} \left( \frac{t_{2ij} - \delta }{\phi}  - d_i - \tau_i - t_{1ij} \right) \beta_{s_{ij}} \mathcal{P}_{\mu_{2s_{ij}}, \sigma_{2s_{ij}}} \left( t_{4ij} - d_i + \frac{\delta - t_{3ij} }{\phi} \right) \right] \nonumber \\
& + (1 - z_{ij}) \ln\left[(1 - \pi_i) \alpha_{r_{ij}} \mathcal{P}_{\mu_{1r_{ij}}, \sigma_{1r_{ij}}} \left( \frac{t_{2ij} - \delta }{\phi}  - d_i - t_{1ij} \right) \beta_{s_{ij}} \mathcal{P}_{\mu_{2s_{ij}}, \sigma_{2s_{ij}}} \left( t_{4ij} - d_i + \frac{\delta - t_{3ij} }{\phi} \right)\right]  \nonumber \\
&  - 2NP\ln\phi + \sum_{i=1}^N \sum_{j=1}^{P_t} \ln\left[ \alpha_{\tilde{r}_{ij}} \mathcal{P}_{\mu_{1\tilde{r}_{ij}}, \sigma_{1\tilde{r}_{ij}}} \left( \tilde{w}_{1ij} \right) \right.  \left.\beta_{\tilde{s}_{ij}} \mathcal{P}_{\mu_{2\tilde{s}_{ij}}, \sigma_{2\tilde{s}_{ij}}} \left( \tilde{w}_{2ij} \right) \right]. \label{Complete_logLL}
\end{align}

We now describe the steps of the EM algorithm. The E-step of the EM algorithm performs an average over the unavailable parts of the complete data conditioned on the incomplete data and current parameter estimates $\hat{\bm{\Omega}}_{\pi}^{'}$ as in $Q(\bm{\Omega}_{\pi}|\hat{\bm{\Omega}}_{\pi}^{'})  =  E\left\{  \mathcal{L}_{com}({\bm{\Omega}}_{\pi}|\bm{X}_c) \bigg | \bm{X}, \hat{\bm{\Omega}}_{\pi}^{'} \right\}$
\begin{align}
 = & \sum_{i=1}^N \sum_{j=1}^P \sum_{k=1}^{M_i} \sum_{l=1}^{L_i} \chi^{(1)}_{ijkl}\ln \bigg[\pi_i\alpha_{ik} \mathcal{P}_{\mu_{1ik}, \sigma_{1ik}} \left( \frac{t_{2ij} - \delta }{\phi}  - d_i - \tau_i - t_{1ij} \right)  \left.\beta_{il} \mathcal{P}_{\mu_{2il}, \sigma_{2il}} \left( t_{4ij} - d_i + \frac{\delta - t_{3ij} }{\phi} \right) \right] \nonumber \\
& + \chi^{(0)}_{ijkl} \ln\left[(1 - \pi_i) \alpha_{ik}\mathcal{P}_{\mu_{1ik}, \sigma_{1ik}} \left( \frac{t_{2ij} - \delta }{\phi}  - d_i - t_{1ij} \right) \right.  \left. \beta_{il} \mathcal{P}_{\mu_{2il}, \sigma_{2il}} \left( t_{4ij} - d_i + \frac{\delta - t_{3ij} }{\phi} \right)\right]  \nonumber \\
& + \sum_{i=1}^N \sum_{j=1}^{P_t} \sum_{k=1}^{M_i} \sum_{l=1}^{L_i} \tilde{a}_{ijkl} \ln \Big[ \alpha_{ik} \mathcal{P}_{\mu_{1ik}, \sigma_{1ik}} \left( \tilde{w}_{1ij} \right)  \beta_{il} \mathcal{P}_{\mu_{2il}, \sigma_{2il}} \left( \tilde{w}_{2ij} \right)\Big] - 2NP\ln\phi, \label{Q_logLL}
\end{align}

where $\chi_{ijkl}^{(1)}, \chi_{ijkl}^{(0)}, \tilde{a}_{ijkl}$ are defined in (\ref{a1_ijkl_men1}), (\ref{a0_ijkl_men1}) and (\ref{a_noise_ijkl_men1}), respectively. The EM algorithm \cite{Blimes} {\color{black} updates} the parameter estimates $\hat{\bm{\Omega}}_{\pi}^{'}$ to new values $\hat{\bm{\Omega}}_{\pi}$ that maximize $Q(\bm{\Omega}_{\pi}|\hat{\bm{\Omega}}_{\pi}^{'})$ in (\ref{Q_logLL}). This is called the M-step of the EM algorithm, and the updated parameters are guaranteed to not decrease the incomplete data likelihood, defined in (\ref{Incomplete_logLL}). {\color{black} The SAGE algorithm inherits this property.} The parameter set $\bm{\Omega}_{\pi}$ is partitioned into the following subsets: $\bm{\Omega}_{\pi, 1} = \{\bm{\pi}, \bm{\alpha}_1, \cdots, \bm{\alpha}_N, \bm{\beta}_1, \cdots, \bm{\beta}_N  \}$, $\bm{\Omega}_{\pi, 2} = \{\bm{\mu}_{11}, \cdots, \bm{\mu}_{1N}, \bm{\mu}_{21}, \cdots, \bm{\mu}_{2N}  \}$, $\bm{\Omega}_{\pi, 3} = \{\bm{\sigma}_{11}, \cdots, \bm{\sigma}_{1N}, \bm{\sigma}_{21}, \cdots, \bm{\sigma}_{2N}  \}$, $\bm{\Omega}_{\pi, 4} = \{d_1, d_2, \cdots, d_N  \}$, $\bm{\Omega}_{\pi, 5} = \{\tau_1, \tau_2, \cdots, \tau_N  \}$, $\bm{\Omega}_{\pi, 6} = \delta$, $\bm{\Omega}_{\pi, 7}  = \phi$. First, $Q(\bm{\Omega}_{\pi}|\hat{\bm{\Omega}}_{\pi}^{'})$ is first maximized with respect to $\bm{\Omega}_{\pi, 1}$ with all other parameters fixed at the current parameter estimates. Then, $\bm{\Omega}_{\pi, 1}$ is set equal to the updated parameter estimate, after which, $Q(\bm{\Omega}_{\pi}|\hat{\bm{\Omega}}_{\pi}^{'})$ is maximized with respect to $\bm{\Omega}_{\pi, 2}$ with all other parameters fixed at the current parameter estimates. This procedure is repeated for all the parameter subsets $\bm{\Omega}_{\pi, 1}, \bm{\Omega}_{\pi, 2}, \cdots, \bm{\Omega}_{\pi, 7}$ until the algorithm converges {\color{black} (small change in $Q(\bm{\Omega}_{\pi}|\hat{\bm{\Omega}}_{\pi}^{'})$)}. 
}

\section{Initialization of parameters for SAGE algorithm}\label{App_sec3}
As the objective function for the optimization problem in (\ref{main_equation_EM}) is not necessarily convex, proper initialization of the various parameters is employed to promote convergence to the global minimum instead of local minimums. We present a simple ad-hoc scheme to obtain the initial values of the various parameters, denoted by $\hat{\bm{\Omega}}_{\pi}^{(0)}$ for the SAGE algorithm. The steps of the initialization are enumerated below:

\begin{algorithmic}[1]
	
	\STATE First, we define new variables $\gamma_i = (\phi(d_i + \tau_i) + \delta)$ and $\zeta_i = (-\phi d_i + \delta)$ for $i = 1, 2, \cdots, N$.

	\FOR {$i = 1 : N$}
	
	\STATE For a given value of $M_i$, run the EM algorithm for the GMM using the update equations given in \cite{Blimes} on  $\tilde{\bm{w}}_{1i}$ to obtain $\hat{\alpha}_{ik}^{(0)}$, $\hat{\mu}_{1ik}^{(0)}$ and $\hat{\sigma}_{1ik}^{(0)}$.
	
	\STATE For a given value of $L_i$, run the EM algorithm for the GMM using the update equations given in \cite{Blimes} on  $\tilde{\bm{w}}_{2i}$ to obtain $\hat{\beta}_{il}^{(0)}$, $\hat{\mu}_{2il}^{(0)}$ and $\hat{\sigma}_{2il}^{(0)}$.
	
	\STATE Using $\hat{\alpha}_{ik}^{(0)}$, $\hat{\mu}_{1ik}^{(0)}$ and $\hat{\sigma}_{1ik}^{(0)}$,  construct an approximate pdf for $f_{1i}(.)$, denoted by $\tilde{f}_{1i}(.)$. Similarly, using $\hat{\beta}_{il}^{(0)}$, $\hat{\mu}_{2il}^{(0)}$ and $\hat{\sigma}_{2il}^{(0)}$, construct an approximate pdf for $f_{2i}(.)$, denoted by $\tilde{f}_{2i}(.)$.
	
	\STATE Consider the timestamps $\bm{t}_{1i}$ and $\bm{t}_{2i}$ from the $i^{\mbox{th}}$ master-slave communication path. We know\footnote{Since we do not have  prior information  on whether the $i^{\mbox{th}}$ path has an unknown asymmetry, we assume $\tau_i \ne 0$.} $\bm{t}_{2i} = (\bm{t}_{1i} + \bm{w}_{1i})\phi + \gamma_i\mathds{1}_P$.	Relaxing the dependency {\color{black} of $\gamma_i$ on $\phi$,} we use the optimum CSOE scheme proposed in \cite{Karthik_Skew_Offset} to obtain an estimate of $\phi$ and $\gamma_i$, denoted by $\hat{\phi}_{fwd, i}$ and $\hat{\gamma}_i$ respectively.

	\STATE Consider the timestamps $\bm{t}_{3i}$ and $\bm{t}_{4i}$ from the $i^{\mbox{th}}$ master-slave communication path. We have $\bm{t}_{3i} = (\bm{t}_{4i} - \bm{w}_{2i})\phi + \zeta_i\mathds{1}_P$. Relaxing the dependency {\color{black} of $\zeta_i$ on $\phi$}, we use the optimum CSOE scheme in \cite{Karthik_Skew_Offset}  to obtain an estimate of $\phi$ and $\zeta_i$, denoted by $\hat{\phi}_{rev, i}$ and $\hat{\zeta}_i$ respectively.
	
	\STATE We then construct the estimate of the clock skew  from the timestamps exchanged in the $i^{\mbox{th}}$ master-slave communication path, denoted by $\hat{\phi}_i$, as $\hat{\phi}_{i}  =  (\hat{\phi}_{fwd, i} + \hat{\phi}_{rev, i})/2$. Similarly, we calculate an estimate of $\delta$, denoted by $\hat{\delta}_{i}$, from the timestamps exchanged in the $i^{\mbox{th}}$ master-slave communication path  as $\hat{\delta}_{i} =  (\hat{\gamma}_i + \hat{\zeta}_i)/2$.

	\ENDFOR

	\STATE Using the obtained estimates $\hat{\phi}_i$ and $\hat{\delta}_i$ for $i = 1, 2, \cdots, N$, we fix our initial estimate of $\delta$, denoted by $\hat{\delta}^{(0)}$, as $\hat{\delta}^{(0)} =  \mbox{median}\{\hat{\delta}_1, \hat{\delta}_2, \cdots, \hat{\delta}_N    \}$. Similarly, we fix our initial estimate of $\phi$, denoted by $\hat{\phi}^{(0)}$, as $\hat{\phi}^{(0)}  = \mbox{mean}\{\hat{\phi}_1, \hat{\phi}_2, \cdots, \hat{\phi}_N \}$.

	\FOR {$i = 1 : N$}
	
	\STATE Estimate the total deterministic path delay in the forward path of the $i^{\mbox{th}}$ master-slave communication path as $\hat{d}_{fwd, i} = (\hat{\gamma}_i - \hat{\delta}^{(0)})/\hat{\phi}^{(0)}$. Similarly, estimate the total deterministic path delay in the reverse path of the $i^{\mbox{th}}$ master-slave communication path as $\hat{d}_{rev, i} =  (\hat{\delta}^{(0)} - \hat{\zeta}_i)/\hat{\phi}^{(0)}$.

	\STATE Set $\hat{\pi}_i^{(0)}$ as $\frac{e^{\left(|\hat{d}_{fwd, i} - \hat{d}_{rev, i}| - d_{\tau}\right)\kappa}}{e^{\left(|\hat{d}_{fwd, i} - \hat{d}_{rev, i}| - d_{\tau}\right)\kappa} + 1}$, where $\kappa$ is a normalization constant\footnote{\color{black} We use a softmax function and assume the delays are in microseconds. So $\kappa = 10^6$. The parameter $\kappa$ can be modified in other scenarios.}.

	\IF {$\left|\hat{d}_{fwd, i} - \hat{d}_{rev, i}\right| \le d_\tau$}
	\STATE Set $\hat{d}_i^{(0)}$ to to $(\hat{d}_{fwd, i} + \hat{d}_{rev, i})/2$ and $\hat{\tau}_i^{(0)}$ to $0$.
	\ELSE
	\STATE Set $\hat{d}_i^{(0)}$ to to $\hat{d}_{rev, i}$ and $\hat{\tau}_i^{(0)}$ to $(\hat{d}_{fwd, i} - \hat{d}_{rev, i})$.
	\ENDIF
	\ENDFOR
\end{algorithmic}


%





%
%

\ifCLASSOPTIONcaptionsoff
  \newpage
\fi



\bibliographystyle{IEEEtran}
\bibliography{IEEEabrv,refs}

\end{document}